\documentclass[12pt,english]{article}
\usepackage{a4wide}
\newcommand{\case}[2]{{\scriptstyle \frac{#1}{#2}}}

\usepackage[T1]{fontenc}
\usepackage[latin1]{inputenc}
\usepackage{amsmath}
\usepackage{graphicx}
\usepackage{amssymb}
\usepackage{color}
\usepackage{subfigure}
\usepackage{rotating}
\usepackage{bm}
\usepackage{axodraw}

\usepackage{amsmath,amssymb}

\newcommand{\E}{\mathrm{e}}
\newcommand{\I}{\mathrm{i}}
\newcommand{\be}{\begin{eqnarray}}
\newcommand{\ee}{\end{eqnarray}}

\newcommand{\nn}{\nonumber }
  
\newcommand{\trc}{\text{tr}_{\text{c}}}

\newcommand{\Li}{\mathrm{Li}}
\newcommand{\Gk}{\Gamma_k}
\newcommand{\Tc}{T_{\text{cr}}}
\newcommand{\yb}{\bar\psi}
\newcommand{\xsb}{$\chi$SB}
\newcommand{\Nf}{N_{\text{f}}}
\newcommand{\Nc}{N_{\text{c}}}
\newcommand{\pat}{\partial_t}

\newcommand{\hath}[1]{#1}
\newcommand{\lp}{\hath{\lambda}_{+}}
\newcommand{\lm}{\hath{\lambda}_{-}}

\newcommand{\lsf}{\hath{\lambda}_{\sigma}}
\newcommand{\lva}{\hath{\lambda}_{\text{VA}}}
\newcommand{\lF}{l_1^{\text{(F)}}}
\newcommand{\lFB}{l^{\textrm{(FB)}}_{1,2}}
\newcommand{\lFBo}{l^{\textrm{(FB)}}_{1,1}}
\newcommand{\Cas}{C_2(\Nc)}

\def\slash#1{\setbox0=\hbox{$#1$}               
   \dimen0=\wd0                                 
   \setbox1=\hbox{/} \dimen1=\wd1               
   \ifdim\dimen0>\dimen1                        
      \rlap{\hbox to \dimen0{\hfil/\hfil}}      
      #1                                        
   \else                            
      \rlap{\hbox to \dimen1{\hfil$#1$\hfil}}   
      /                                         
   \fi}                                         %

\usepackage{babel}
\makeatother
\begin{document}

\centerline{\Large\bf Chiral phase boundary of QCD at finite
temperature}

\vspace{.8cm}

\centerline{Jens Braun and Holger Gies}

\vspace{0.6cm}

\centerline{\small\it Institut f\"ur Theoretische Physik,
  Philosophenweg 16 and 19, 69120 Heidelberg, Germany}

\centerline{\small\it  E-mail:  jbraun@tphys.uni-heidelberg.de,
  h.gies@thphys.uni-heidelberg.de} 

\begin{abstract}
  We analyze the approach to chiral symmetry breaking in QCD at finite
  temperature, using the functional renormalization group. We compute
  the running gauge coupling in QCD for all temperatures and scales
  within a simple truncated renormalization flow. At finite
  temperature, the coupling is governed by a fixed point of the
  3-dimensional theory for scales smaller than the corresponding
  temperature. Chiral symmetry breaking is approached if the running
  coupling drives the quark sector to criticality. We quantitatively
  determine the phase boundary in the plane of temperature and number
  of flavors and find good agreement with lattice results. As a
  generic and testable prediction, we observe that our underlying IR
  fixed-point scenario leaves its imprint in the shape of the phase
  boundary near the critical flavor number: here, the scaling of the
  critical temperature is determined by the zero-temperature IR
  critical exponent of the running coupling.
\end{abstract}




\section{Introduction and summary}

The properties of strongly interacting matter change distinctly during
the transition from low to high temperatures \cite{Karsch:2003jg}, as
is currently explored at heavy-ion colliders. Whereas the
low-temperature phase can be described in terms of ordinary hadronic
states, a copious excitation of resonances in a hot hadronic gas
eventually implies the breakdown of the hadronic picture; instead, a
description in terms of quarks and gluons is expected to arise
naturally owing to asymptotic freedom. In the transition region
between these asymptotic descriptions, effective degrees of freedom,
such as order parameters for the chiral or deconfining phase
transition, may characterize the physical properties in simple terms,
i.e., with a simple effective action \cite{Pisarski:1983ms}.

Recently, the notion of a strongly interacting high-temperature plasma
phase has attracted much attention \cite{Shuryak:2003xe}, implying
that any generic choice of degrees of freedom will not lead to a
weakly coupled description. In fact, it is natural to expect
  that the low-energy modes of the thermal spectrum still remain
  strongly coupled even above the phase transition. If so, a
formulation with microscopic degrees of freedom from first principles
should serve as the most powerful and flexible approach to a
quantitative understanding of the system for a wide parameter range.

In this microscopic formulation, an expansion in the coupling constant
is a natural first step \cite{Shuryak:1977ut}. The structure of this
expansion turns out to be theoretically involved
\cite{Kapusta:1979fh}, exhibiting a slow convergence behavior
\cite{Arnold:1994ps} and requiring coefficients of nonperturbative
origin \cite{Linde:1980ts}. Still, a physically well-understood
computational scheme can be constructed with the aid of
effective-field theory methods \cite{Braaten:1991gm}. This facilitates
a systematic determination of expansion coefficients, and the
agreement with lattice simulations is often surprisingly good down to
temperatures close to $T_{\text{cr}}$ \cite{Laine:2003ay}.  The
phase-transition region and the deep IR, however, remain inaccessible
with such an expansion.

In the present work, we use a different expansion scheme to study
finite-temperature Yang-Mills theory and QCD in terms of microscopic
variables, i.e., gluons and quarks. This scheme is based on a
systematic and consistent operator expansion of the effective action
which is inherently nonperturbative in the coupling.  For bridging the
scales from weak to strong coupling, we use the functional
renormalization group (RG) \cite{Wegner,Wetterich:yh,bonini} which is
particularly powerful for analyzing phase transitions and critical
phenomena. 

Since we do not expect that microscopic variables can answer all
relevant questions in a simple fashion, we concentrate on two
accessible problems. In the first part, we focus on the running of the
gauge coupling driven by quantum as well as thermal fluctuations of
pure gluodynamics. Our findings generalize similar previous
zero-temperature studies to arbitrary values of the temperature
\cite{Gies:2002af}. In the second part, we employ this result for an
investigation of the induced quark dynamics including its
back-reactions on gluodynamics, in order to monitor the status of
chiral symmetry at finite temperature. This strategy facilitates a
computation of the critical temperature above which chiral symmetry is
restored. Generalizing the system to an arbitrary number of quark
flavors, we explore the phase boundary in the plane of temperature and
flavor number. First results of our investigation have already been
presented in \cite{Braun:2005uj}. In the present work, we detail our
approach and generalize our findings. We also report on results for
the gauge group SU(2), develop the formalism further for finite quark
masses, and perform a stability analysis of our results.  Moreover, we
gain a simple analytical understanding of one of our most important
results: the shape of the chiral phase boundary in the
$(T,\Nf)$ plane. Whereas fermionic screening is the dominating
mechanism for small $\Nf$, we observe an intriguing relation between
the $\Nf$ scaling of the critical temperature near the critical flavor
number and the zero-temperature IR critical exponent of the running
gauge coupling. This relation connects two different universal
quantities with each other and, thus, represents a generic testable
prediction of the phase-transition scenario, arising from our
truncated RG flow.

In Sect.~\ref{sec:RG_gauge}, we summarize the technique of RG flow
equations in the background-field gauge, which we use for the
construction of a gauge-invariant flow. In Sect.~\ref{sec:RunCoup}, we
discuss the details of our truncation in the gluonic sector and
evaluate the running gauge coupling at zero and finite
temperature. Quark degrees of freedom are included in
Sect.~\ref{sec:quarks} and the general mechanisms of chiral quark
dynamics supported by our truncated RG flow is elucidated. Our
findings for the chiral phase transition are presented in
Sect.~\ref{sec:xsb}, our conclusions and a critical assessment of our
results are given in Sect.~\ref{sec:conc}.

\section{RG flow equation in background-field gauge}
\label{sec:RG_gauge}

As an alternative to the functional-integral definition of quantum
field theory, we use a differential formulation provided by the
functional RG \cite{Wegner,Wetterich:yh,bonini}. In this approach,
flow equations for general correlation functions can be constructed
\cite{Pawlowski:2005xe}. A convenient version is given by the flow
equation for the effective average action $\Gamma_k$ which
interpolates between the bare action $\Gamma_{k=\Lambda}= S$ and the
full quantum effective action $\Gamma=\Gamma_{k=0}$
\cite{Wetterich:yh}. The latter corresponds to the generator of
fully-dressed proper vertices. Aiming at gluodynamics, a
gauge-invariant flow can be constructed with the aid of the
background-field formalism \cite{Abbott:1980hw}, yielding the flow
equation \cite{Reuter:1993kw}
\begin{equation}
k\,\partial_{k}\Gamma_{k}[A,\bar{A}]
\equiv\partial_{t}\Gamma_{k}[A,\bar{A}]
=\frac{1}{2}\mathrm{STr}{\displaystyle
\frac{\partial_{t}R_{k}(\Gamma_{k}^{(2)}[\bar{A},\bar{A}])}
{\Gamma_{{\scriptstyle k}}^{(2)}[A,\bar{A}]
+R_{k}(\Gamma_{k}^{(2)}[\bar{A},\bar{A}])}}
, \quad t=\ln\frac{
k}{\Lambda}.
\label{eq:flow_eq1}
\end{equation}  
Here, $\Gamma_{k}^{(2)}$ denotes the second functional derivative with
respect to the fluctuating field $A$, whereas the background-field
denoted by $\bar{A}$ remains purely classical. The ghost fields
  are not displayed here and in the following for brevity, but the
  super-trace also includes a trace over the ghost sector with the
  corresponding minus sign.  The regulator $R_{k}$ in the denominator
suppresses infrared (IR) modes below the scale $k$, and its derivative
$k\partial_{k}R_{k}$ ensures ultraviolet (UV) finiteness; as a
consequence, the flow of $\Gk$ is dominated by fluctuations with
momenta $p^2\simeq k^2$, implementing the concept of smooth
momentum-shell integrations.

The background-field formalism allows for a convenient definition of a
gauge-invariant effective action obtained by a gauge-fixed calculation
\cite{Abbott:1980hw}. For this, an auxiliary symmetry in the form of
gauge-like transformations of the background field $\bar A$ is
constructed which remains manifestly preserved during the
calculation. Identifying the background field with the expectation
value $A$ of the fluctuating field at the end of the calculation,
$A=\bar A$, the quantum effective action $\Gamma$ inherits the
symmetry properties of the background field and thus is gauge
invariant, $\Gamma[A]=\Gamma[A,\bar A =A]$.

The background-field method for flow equations has been presented in
\cite{Reuter:1993kw}: the gauge fixing together with the
regularization lead to gauge constraints for the effective action,
resulting in regulator-modified Ward-Takahashi identities
\cite{Reuter:1997gx,Freire:2000bq}, see also
\cite{Ellwanger:iz,Pawlowski:2005xe}. In this work, we solve the flow
approximately, following the strategy developed in
\cite{Reuter:1997gx,Gies:2002af}. The property of manifest gauge
invariance of the solution is still maintained by the approximation of
setting $A=\bar A$ already for finite values of $k$. Thereby, we
neglect the difference between the RG flows of the fluctuating and the
background field (see \cite{Pawlowski:2001df} for a treatment of this
difference).  The price to be paid for this approximation is that the
flow is no longer {\em closed} \cite{Litim:2002xm}; i.e., information
required for the next RG step is not completely provided by the
preceding step. Moreover, this approximation satisfies some but not
all constraints imposed by the regulator-modified Ward-Takahashi
identities (mWTI).  Here we assume that both the information loss and
the corrections due to the mWTI are quantitatively negligible for the
final result. The advantage of the approximation of using
$\Gamma_{k}[A,\bar A =A]$ for all $k$ is that we obtain a
gauge-invariant approximate solution of the quantum
theory. \footnote{For recent advances of an alternative approach which
is based on a manifestly gauge invariant regulator, see
\cite{Morris:2000fs}. Further proposals for thermal gauge-invariant
flows can be found in \cite{D'Attanasio:1996fy}.}

In the present work, we optimize our truncated flow by inserting the
background-field dependent $\Gamma^{(2)}$ into the regulator in
Eq.~\eqref{eq:flow_eq1}. This adjusts the regularization to the
spectral flow of the fluctuations \cite{Gies:2002af,Litim:2002xm}; it
also implies a significant improvement, since larger classes of
diagrams can be resummed in the present truncation scheme.  As another
advantage, the background-field method together with the
identification $A=\bar A$ for all $k$ allows to bring the flow
equation into a propertime form
\cite{Gies:2002af,Litim:2002xm,Litim:2002hj} which generalizes
standard propertime flows \cite{Liao:1994fp}; the latter have often
successfully be used for low-energy QCD models \cite{Schaefer:em}. For
this, we use a regulator $R_k$ of the form
\begin{equation}
R_{k}(x)=xr(y),\quad
y:=\frac{x}{\mathcal{Z}_{k}k^{2}}\,,\label{eq:reg_def}
\end{equation}
with $r(y)$ being a dimensionless regulator shape function of
dimensionless argument. Here $\mathcal{Z}_{k}$ denotes a wave-function
renormalization.  Note that both $R_{k}$ and $\mathcal{Z}_{k}$ are
matrix-valued in field space. A natural choice for the matrix entries
of $\mathcal{Z}_k$ is given by the wave function renormalizations of
the corresponding fields, since this establishes manifest RG
invariance of the flow equation.\footnote{For the longitudinal gluon
components, this implies that the matrix entry
$(\mathcal{Z}_k)_{\text{LL}}$ is proportional to the inverse
gauge-fixing parameter $\xi$. As a result, this renders the truncated
flow independent of $\xi$, and we can implicitly choose the Landau
gauge $\xi\equiv0$ which is known to be an RG fixed point
\cite{Ellwanger:1995qf,Litim:1998qi}.}  More properties of the
regulator are summarized in Appendix
\ref{sec:Regulator_function}. Identifying the background field and the
fluctuation field, the flow equation yields
\begin{equation}
\partial_{t}\Gamma_{k}[A\!
  =\!\bar{A},\bar{A}]=\frac{1}{2}\mathrm{STr}
  \partial_{t}R_{k}(\Gamma_{k}^{(2)})[\Gamma_{{\scriptstyle
        k}}^{(2)}+R_{k}]^{-1}
=\frac{1}{2}{\displaystyle
    \int_{0}^{\infty}ds\,\mathrm{STr}\hat{f}(s,\eta_{\mathcal Z})
\exp\big(-\frac{s}{k^{2}}\Gamma_{{\scriptstyle k}}^{(2)}\big)\,.}
\label{eq:fe_PT} 
\end{equation}
Here, we have introduced the (matrix-valued) anomalous
dimension
\begin{equation}
\eta_{\mathcal Z}:=-\partial_{t}\ln\mathcal{Z}_{k}
  =-\frac{1}{\mathcal{Z}_{k}}\partial_{t}\mathcal{Z}_{k}.\label{eq:defeta}
\end{equation}
The operator $\hat{f}(s,\eta_{\mathcal Z})$ represents the translation
of the regulator $R_k$ into propertime space given by
\begin{equation}
\hat{f}(s,\eta_{\mathcal Z}) =\tilde{g}(s)(2-\eta_{\mathcal Z})
   +(\tilde{H}(s)-\tilde{G}(s))\frac{1}{s}\partial_{t}.
\label{eq:fop}
\end{equation}
The auxiliary functions on the RHS are related to the regulator shape
function $r(y)$ by Laplace transformation:
\begin{eqnarray}
h(y)&=&\frac{-yr'(y)}{1+r(y)},\quad
h(y)=\int_{0}^{\infty}ds\,\tilde{h}(s)\E^{-ys},\quad
\frac{d}{ds}\tilde{H}(s)=\tilde{h}(s)\,,\quad\tilde{H}(0)=0,
\label{eq:def_h}\\
g(y)&=&\frac{r(y)}{1+r(y)},\quad 
g(y)=\int_{0}^{\infty}ds\,\tilde{g}(s)\E^{-ys},\quad
\frac{d}{ds}\tilde{G}(s)=\tilde{g}(s),\quad\tilde{G}(0)=0.
\label{eq:def_g}
\end{eqnarray}
So far, we have discussed pure gauge theory. Quark fields with a
mass matrix $M_{\bar{\psi}\psi}$ can similarly be treated within our
framework. For this, we use a regulator $R_{k} ^{\psi}$ of the
form \cite{Gies:2004hy}
\begin{equation}
R_{k}^{\psi}(\I\slash{\bar{D}})=Z_\psi\I\slash{\bar{D}}\, r_{\psi}
\Big({\textstyle \frac{(\I\slash{\bar{D}})^{2}}{k^{2}}}\Big)\,,
\label{eq:qu_reg}
\end{equation}
where $\slash{\bar{D}}$ is a short-hand notation for
$\slash{\partial}-i\bar{g}\slash{\bar{A}}$.  Note that the quark
fields live in the fundamental representation. This form of the
fermionic regulator is chirally symmetric as well as invariant under
background-field transformations.  For later purposes, let us list
the quark-fluctuation contributions to the gluonic sector; the flow
of $\Gamma_k[\bar A]$ induced by quarks with the regulator
\eqref{eq:qu_reg} can also be written in propertime form,
\begin{equation}
\pat\Gamma _{k}[\bar{A}]\big|_{\psi}=-\mathrm{Tr}
 {\pat R_{k}^{\psi}(\I\slash{\bar{D}})} 
[{\Gamma_{k}^{(2)}+R_{k}}]^{-1}_{\psi}
=-\int_{0}^{\infty}ds\,\mathrm{Tr}\hat{f}_{\psi}(s,\eta_{\psi},
{\textstyle\frac{M_{\bar{\psi}\psi}}{k}}) 
\exp\big(-\frac{s}{k^{2}}(\I\slash{\bar{D}})^{2}\big)\,,
\label{eq:fe_quark} 
\end{equation}
with $[\Gamma _{k}^{(2)}+R_k]^{-1}_\psi$ denoting the exact
(regularized) quark propagator in the background field.
In Eq. \eqref{eq:fe_quark}, we have introduced the
anomalous dimension of the quark field,
\begin{equation}
\eta_{\psi}:=-\pat\ln{Z}_{\psi}
 .\label{eq:eta_def_quark} 
\end{equation}
In complete analogy to the gauge sector, we define the
operator $\hat{f}_{\psi}(s,\eta_{\psi},\tilde{m})$ by
\begin{equation}
\hat{f}_{\psi}(s,\eta_{\psi},\tilde{m})=\tilde{g}^{\psi}(s,\tilde{m})(1-\eta_{\psi}) 
   +(\tilde{H}^{\psi}(s,\tilde{m})-\tilde{G}^{\psi}(s,\tilde{m}))\frac{1}{2s}\partial_{t}\,.
\label{eq:f_psi_op}
\end{equation}
The regulator shape function $r_{\psi}(y)$ is related to the auxiliary
functions appearing in the definition of the operator
$\hat{f}_{\psi}(s,\eta_{\psi},\tilde{m})$ as follows
\begin{eqnarray}
h^{\psi}(y,\tilde{m})
 &=&\frac{-2y^{2}r'_{\psi}(1+r_{\psi})}{y(1+r_{\psi})^{2}
      +\tilde{m}^2}\,,\quad
  g^{\psi}(y,\tilde{m})
    =\frac{yr_{\psi}(1+r_{\psi})}{y(1+r_{\psi})^{2}+\tilde{m}^2}\,, 
    \label{eq:def_hpsi}\\
h^{\psi}(y,\tilde{m})
 &=&\int_{0}^{\infty}ds\,\tilde{h}^{\psi}(s,\tilde{m})\E^{-ys}\, 
     \quad
  \frac{d}{ds}\tilde{H}^{\psi}(s,\tilde{m})
    =\tilde{h}^{\psi}(s,\tilde{m}),\quad
  \tilde{H}^{\psi}(0,\tilde{m})=0.
    \label{eq:hprop}
\end{eqnarray}
The corresponding functions $g^{\psi}(y,\tilde{m})$,
$\tilde{g}^{\psi}(s,\tilde{m})$, and $\tilde{G}^{\psi}(s,\tilde{M})$ are
related to each other analogously to Eq.~\eqref{eq:hprop}.
The present construction facilitates a simple inclusion of
finite quark masses without complicating the convenient
(generalized) propertime form of the flow equation.
  
To summarize: the functional traces in Eq.~\eqref{eq:fe_PT} and
\eqref{eq:fe_quark} can now be evaluated, for instance, with powerful
heat-kernel techniques, and all details of the regularization are
encoded in the auxiliary functions $h,g$ etc.  Equations
\eqref{eq:fe_PT} and \eqref{eq:fe_quark} now serve as the starting
point for our investigation of the gluon sector. The flow of
quark-field dependent parts of the effective action proceeds in a
standard fashion \cite{Ellwanger:1994wy}, see \cite{Aoki:2000wm} for
reviews; in particular, a propertime representation is not needed for
the truncation in the quark sector described below.

\section{RG flow of the running coupling at finite temperature}
\label{sec:RunCoup}

At first sight, the running coupling does not seem to be a useful
quantity in the nonperturbative domain, since it is RG-scheme and
strongly definition dependent. Therefore, we cannot a priori associate
a universal meaning to the coupling flow, but have to use and
interpret it always in the light of its definition and RG scheme.

In fact, the background-field formalism provides for a simple
nonperturbative definition of the running coupling in terms of the
background-field wave function renormalization $Z_k$. This is based on
the nonrenormalization property of the product of coupling and
background gauge field, $\bar g \bar A$ \cite{Abbott:1980hw}. The
running-coupling $\beta_{g^2}$ function is thus related to the
anomalous dimension of the background field
(cf. Eq.~\eqref{eq:sec2_2_defcoup} below),
\begin{equation}
\beta_{g^{2}}\equiv\partial_{t}g^{2}=(d-4+\eta)g^{2}, \quad
\eta=-\frac{1}{Z_k} \pat Z_k, \label{betadef}
\end{equation}
where we have kept the spacetime dimension $d$ arbitrary. Since the
background field can naturally be associated with the vacuum of
gluodynamics, we may interpret our coupling as the response strength
of the vacuum to color-charged perturbations.

\subsection{Truncated RG flow}

Owing to strong coupling, we cannot expect that low-energy
gluodynamics can be described by a small number of gluonic
operators. On the contrary, infinitely many operators become RG
relevant and will in turn drive the running of the coupling.
Following the strategy developed in \cite{Reuter:1997gx},
we span a truncated space of effective action functionals by the
ansatz
\begin{equation}
\Gamma_{k}=\Gamma^{\text{YM}}_{k}[A,\bar{A}]
  +\Gamma_{k}^{\text{gf}}[A,\bar{A}]
  +\Gamma_{k}^{\text{gh}}[A,\bar{A},\bar c, c]
  +\Gamma_k^{\text{quark}}[A,\bar A,\yb,\psi]
.\label{gentrunc}
\end{equation} 
Here, $\Gamma^{\text{gf}}$ and $\Gamma^{\text{gh}}$ represent
generalized gauge-fixing and ghost contributions, which we assume to
be well approximated by their classical form in the present work,
\begin{equation}
\Gamma_{k}^{\text{gf}}[A,\bar{A}]=\frac{1}{2\xi}\int_x 
   (D_{\mu}[\bar{A}](A-\bar{A})_{\mu})^{2},\quad
\Gamma_{k}^{\text{gh}}[A,\bar{A},\bar c,c]=-\!\int_x
\bar{c}D_{\mu}[\bar{A}]D_{\mu}[A]c,\quad D[A]=\partial -\I \bar g A,
\label{ghofix}
\end{equation}
neglecting any non-trivial running in these sectors. Here, $\bar g$
denotes the bare coupling, and the gauge field lives in the adjoint
representation, $A_\mu=A_\mu^c T^c$, with hermitean gauge-group
generators $T^c$. The gluonic part $\Gamma_k^{\text{YM}}$ carries the
desired physical information about the quantum theory that can be
gauge-invariantly extracted in the limit
$\Gamma_{k} ^{\text{YM}}[A]=\Gamma_{k}^{\text{YM}}[A,\bar A~=~A]$.

The quark contributions are contained in
\begin{equation}
\Gamma_k^{\psi}[A,\yb,\psi]
 = \int_x \yb(\I\slash{D}[A]+M_{\yb\psi})\psi +
 \Gamma_k^{\text{q-int}}[\yb,\psi], 
\label{eq:quarktrunc1}
\end{equation}
where $M_{\yb\psi}$ denotes the quark mass matrix, and the quarks
transform under the fundamental representation of the gauge group. The
last term $\Gamma_k^{\text{q-int}}[\yb,\psi]$ denotes our ansatz for
gluon-induced quark self-interactions to be discussed in
Sect.~\ref{sec:quarks}. In Eq. \eqref{eq:quarktrunc1}, we have already
set the quark wave function renormalization to $Z_\psi=1$, which is a
combined consequence of the Landau gauge and our later choice for
$\Gamma_k^{\text{q-int}}[\yb,\psi]$. 

An infinite but still tractable set of gauge-field operators is
given by the nontrivial part of our gluonic truncation,
\begin{equation}
\Gamma_{k}^{\text{YM}}[A]=\int_x \mathcal W_k(\theta), \quad
\theta=\frac{1}{4} 
F_{\mu\nu}^a F_{\mu\nu}^a. \label{Wtrunc}
\end{equation}
Expanding the function $\mathcal W(\theta)=W_1 \theta+ \frac{1}{2}
W_{2} \theta^2+\frac{1}{3!} W_3 \theta^3\dots$, the expansion
coefficients $W_i$ denote an infinite set of generalized couplings.
Here, $W_1$ is identical to the desired background-field wave function
renormalization, $Z_k\equiv W_1$, defining  the running of the
coupling,
\begin{equation}
g^{2} = k^{d-4}{Z}_{k}^{-1}\bar{g}^{2},\label{eq:sec2_2_defcoup}
\end{equation}
which Eq. \eqref{betadef} is a consequence of. This truncation
corresponds to a gradient expansion in the field strength, neglecting
higher-derivative terms and more complicated color and Lorentz
structures. In this way, the truncation includes arbitrarily high
gluonic correlators projected onto their small-momentum limit and onto
the particular color and Lorentz structure arising from powers of
$F^2$.  In our truncation, the running of the coupling is successively
driven by all generalized couplings $W_i$.

It is convenient to express the flow equation in terms of dimensionless
renormalized quantities
\begin{eqnarray}
\vartheta & = & g^{2}k^{-d}{Z}_{k}^{-1}\theta\equiv
   k^{-4}\bar{g}^{2}\theta,\\ 
w(\vartheta) & = & g^{2}k^{-d}\mathcal{W}_{k}(\theta)\equiv 
   k^{-4}{Z}_{k}^{-1}\bar{g}^{2}\mathcal{W}_{k}(k^{4}\vartheta/\bar{g}^{2}).
\label{eq:renq}
\end{eqnarray}
Inserting Eq. \eqref{gentrunc} into Eq. \eqref{eq:fe_PT},
we obtain the flow equation for $w(\vartheta)$:
\begin{multline}
\partial_{t}w=-(4-\eta)w+4\vartheta\dot
  w+\frac{g^{2}}{2(4\pi)^{\frac{d}{2}}}\int_{0}^{\infty}\!
  ds\,\Bigg\{-16\sum_{i=1}^{\Nc}
\sum_{\xi=1}^{\Nf}\tilde{h}^{\psi}(s,{\textstyle
  \frac{m_{\xi}}{k}})f_{T}^{\psi}(s,{\textstyle
  \frac{T}{k}})f^{\psi}(sb_{i})b_{i}^{e_{d}}\\ 
+\tilde{h}(s)\Bigg[4\sum_{l=1}^{\Nc^{2}-1}\!
\Big(f_{T}^{A}(s\!\stackrel{.}{w},{\textstyle
  \frac{T}{k}})f_{1}^{A}(s\!\stackrel{.}{w\!}b_{l})-f_{T}^{A}(s,{\textstyle
  \frac{T}{k}})f_{2}^{A}(sb_{l})\Big)b_{l}^{e_{d}}\\   
-2f_{T}^{A}(s\!\stackrel{.}{w},{\textstyle
  \frac{T}{k}})f_{3}^{A}(s\!\stackrel{.}{w},\frac{\stackrel{.}{w}}
{\stackrel{.}{w}+2\vartheta\stackrel{..}{w}})\Bigg]
-\Big(\eta\tilde{g}(s)+(\tilde{h}(s)-\tilde{g}(s))\Big(\frac{\partial_{t}
\stackrel{.}{w}-4\vartheta\stackrel{..}{w}}{\stackrel{.}{w}}\Big)\Big)\times\\
\times\Bigg[2\sum_{l=1}^{\Nc^{2}-1}
f_{T}^{A}(s\!\stackrel{.}{w},{\textstyle 
  \frac{T}{k}})f_{1}^{A}(s\!\stackrel{.}{w\!}b_{l})b_{l}^{e_{d}}
-f_{T}^{A}(s\!\stackrel{.}{w},{\textstyle 
  \frac{T}{k}})f_{3}^{A}(s\!\stackrel{.}{w},
\frac{\stackrel{.}{w}}{\stackrel{.}{w}+2\vartheta\stackrel{..}{w}})
\Bigg]\\  
-\frac{2(\tilde{h}(s)-\tilde{g}(s))\vartheta}{(\stackrel{.}{w}+2\vartheta\stackrel{..}{w})^{2}}
\Big(\stackrel{..}{w}\!\partial_{t}\!\stackrel{.}{w}
-\stackrel{.}{w\!}\partial_{t}\!\stackrel{..}{w}
+4\!\stackrel{.}{w}\stackrel{..}{w}+4\vartheta(\stackrel{.}{w}
\stackrel{...}{w}-\stackrel{..}{w}^{2})\Big)
f_{4}^{A}(s\!\stackrel{.}{w},{\textstyle \frac{T}{k}})\Big]
\Bigg\}\,,\label{eq:fe_wpot}
\end{multline}
where the auxiliary functions $f$ are defined in
App.~\ref{sec:Regulator_function}, and we have used the abbreviation
$e_{d}=\frac{d-1}{2}$. The ``color magnetic'' field components $b_i$
are defined by $b_i=|\nu_i|\sqrt{2\vartheta}$, where $\nu_i$ denotes
eigenvalues of $(n^aT^a)$ in the fundamental representation;
correspondingly, $b_l$ is equivalently defined for the adjoint
representation.  Furthermore, we have used the short-hand notation
$w\equiv w(\vartheta)$ and dots denote derivatives with respect to
$\vartheta$. In order to extract the flow equation for the running
coupling, we expand the function $w(\vartheta)$ in powers of
$\vartheta$,
\begin{equation}
w(\vartheta)=\sum_{i=0}^{\infty}\frac{w_{i}}{i!}\vartheta^{i}\,,
\quad w_{1}=1. \label{eq:Wexp}
\end{equation}
Note that $w_1$ is fixed to 1 by definition \eqref{eq:renq}.
Inserting this expansion into Eq. \eqref{eq:fe_wpot}, we obtain an
infinite tower of first-order differential equations for the
coefficients $w_{i}$. In the present work, we concentrate on the
running coupling and ignore the full form of the function
$\mathcal{W}$; hence, we set $w_i\to 0$ for $i\geq2$ on the RHS of the
flow equation as a first approximation, but keep track of the flow of
all coefficients $w_i$. The resulting infinite tower of equations is
of the form
\begin{equation}
\pat w_i= X_i(g^2,\eta)+Y_{ij}(g^2) \pat w_j, \label{eq:tower}
\end{equation}
with known functions $X_i,Y_{ij}$, the latter of which obeys
$Y_{ij}=0$ for $j>i+1$. Note that we have not dropped the $w_i$ flows,
$\pat w_i$, which are a consequence of the spectral adjustment of the
flow. This infinite set of equations can iteratively be solved,
yielding the anomalous dimension as an infinite power series of
$g^{2}$ (for technical details, see \cite{Gies:2002af,Gies:2003ic}), 
\begin{equation}
\eta=\sum_{i=0}^{\infty}a_{m}G^{m}\quad\mathrm{{with}}\quad
G\equiv\frac{g^{2}}{2(4\pi)^{d/2}}\,.\label{eq:sec2_2_eta_series} 
\end{equation}
The coefficients $a_m$ can be worked out analytically; they depend on
the gauge group, the number of quark flavors, their masses, the
temperature and the regulator. Equation \eqref{eq:sec2_2_eta_series}
constitutes an asymptotic series, since the coefficients $a_m$ grow at
least factorially. This is no surprise, since the expansion
\eqref{eq:Wexp} induces an expansion of the propertime integrals in
Eq. \eqref{eq:fe_wpot} for which this is a well-understood property
\cite{Dunne:1999uy}. A good approximation of the underlying finite
integral representation of Eq. \eqref{eq:sec2_2_eta_series} can be
deduced from a Borel resummation including only the leading asymptotic
growth of the $a_m$,
\begin{equation}
\eta\simeq\sum_{i=0}^{\infty}a^{\text{l.g.}}_{m}G^{m}\,.
\label{eq:sec2_2_eta_series2}  
\end{equation}
The leading growth coefficients are given by a sum of gluon/ghost and gluon-quark contributions,
\begin{align}
a_{m}^{\text{l.g.}} &
 =4(-2c_{1})^{m-1}\frac{\Gamma(z_{d}+m)\Gamma(m+1)}{\Gamma(z_{d}+1)}
\Big[\bar{h}_{2m-e_{d}}^{A}({\textstyle
 \frac{T}{k}})(d\!-\!2)\frac{2^{2m}-2}{(2m)!}
\tau_{m}^{A}B_{2m}\nonumber
 \\ 
 &
 \qquad\qquad\qquad\quad
-\frac{4}{\Gamma(2m)}\tau_{m}^{A}\bar{h}_{2m-e_{d}}^{A}({\textstyle
 \frac{T}{k}})
+4^{m+1}\frac{B_{2m}}{(2m)!}
\tau_{m}^{\psi}\sum_{i=1}^{\Nf}\bar{h}_{2m-e_{d}}^{\psi}({\textstyle
 \frac{m_{i}}{k}},{\textstyle
 \frac{T}{k}})\Big].\label{eq:sec_2_2_a_m}  
\end{align}
The auxiliary functions $c_{1}$, $c_{2}$, $z_{d}$ and the
moments $\bar{h}_{j}$, $\bar{h}_{j}^{\psi}$ are defined in 
App. \ref{sec:Regulator_function} and \ref{sec:Resummation}.  The
group theoretical factors $\tau_{m}^{A}$ and $\tau_{m}^{\psi}$ are
defined and discussed in App. \ref{sec:color}. The last term in the
second line of Eq. \eqref{eq:sec_2_2_a_m} contains the quark
contributions to the anomalous dimension. The remaining terms are of
gluonic origin. 

The first term in the second line has to be treated with care, since
it arises from the Nielsen-Olesen mode in the propagator
\cite{Nielsen:1978rm} which is unstable in the IR. This mode occurs in
the perturbative evaluation of gradient-expanded effective actions and
signals the instability of chromo fields with large spatial
correlation. At finite temperature, this problem is particularly
severe, since such a mode will strongly be populated by thermal
fluctuations, typically spoiling perturbative computations
\cite{Dittrich:1980nh}. 

From the flow-equation perspective, this does not cause conceptual
problems, since no assumption on large spatial correlations of the
background field is needed, in contrast to the perturbative gradient
expansion.

For an expansion of the flow equation about the (unknown) true
vacuum state, the regulated propagator would be positive definite,
$\Gamma_k^{(2)}+R_k>0$ for $k>0$. Even without knowing the true vacuum
state, it is therefore a viable procedure to include only the positive
part of the spectrum of $\Gamma_k^{(2)}+R_k$ in our truncation, since
it is an exact operation for stable background fields. At zero
temperature, these considerations are redundant, since the
unstable mode merely creates imaginary parts that can easily be
separated from the real coupling flow. At finite temperature, we only
have to remove the unphysical thermal population of this mode which we
do by a $T$-dependent regulator that screens the instability. As an
unambiguous regularization, we include the Nielsen-Olesen mode for all
$k\geq T$ as it is, dropping possible imaginary parts; for $k<T$ we
remove the Nielsen-Olesen mode completely, thereby inhibiting its
thermal excitation. Of course, a smeared regularization of this mode
is also possible, as discussed in App. \ref{sec:regdep}. Therein, the
regularization used here is shown to be a point of ``minimum
sensitivity'' \cite{Stevenson:1981vj} in a whole class of
regulators. This supports our viewpoint that our regularization has
the least contamination of unphysical thermal population of the
Nielsen-Olesen mode.
\begin{figure}[t]
\begin{center}
\includegraphics[%
  clip,
  scale=0.8]{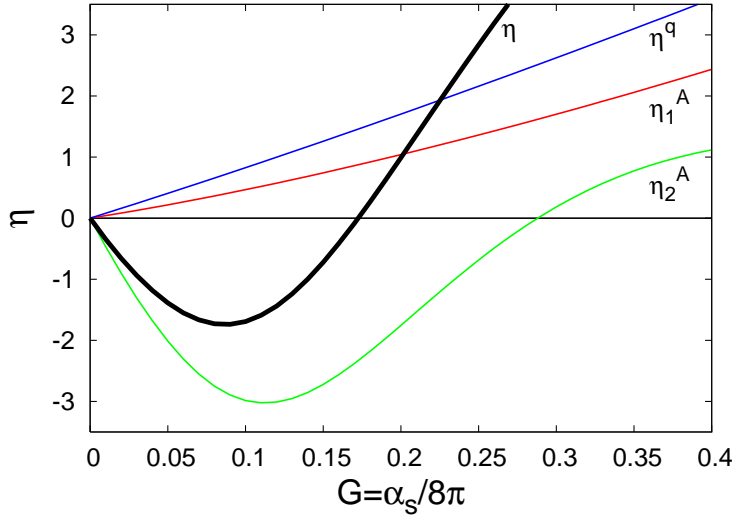}
\end{center}
\caption{Anomalous dimension $\eta$ as a function of
    $G=\frac{\alpha_s}{8\pi}$ for $4d$ $SU(\Nc=3)$ theory with $\Nf=3$
    massless quark flavors at vanishing temperature.  The gluonic
    parts $\eta_{1}^{A},\eta_{2}^{A}$ and the quark part
    $\eta^{\text{q}}$ contributing to the anomalous dimension $\eta$
    (thick black line) are shown separately. The gluonic parts $\eta_1
    ^{A}$ and $\eta_2 ^{A}$ agree with the results found in
    \cite{Gies:2002af}.  The figure shows the results from a
    calculation with a background field pointing into the 8-direction
    in color space.}
\label{fig:etaT0} 
\end{figure}

We outline the resummation of $\eta$ of Eq.
\eqref{eq:sec2_2_eta_series2} in App. \ref{sec:Resummation}, yielding
\begin{equation}
\eta=\eta_{1}^{A}+\eta_{2}^{A}+\eta^{\text{q}}, \label{eq:etares}
\end{equation}
with gluonic parts $\eta_{1}^{A},\eta_{2}^{A}$ and the quark
contribution to the gluon anomalous dimension
$\eta^{\text{q}}$.\footnote{The contribution $\eta^{\text{q}}$
should not be confused with the quark anomalous dimension $\eta_\psi$
which is zero in our truncation.} Finite integral representations of
these functions are given in Eqs.~\eqref{eq:app_etaa2},
\eqref{eq:app_etab2}, and \eqref{eq:eta_quark2}. For pure
gluodynamics, $\eta_{1}^{A}$ and $\eta_{2}^{A}$ carry the full
information about the running coupling.

In Fig. \ref{fig:etaT0}, we show the result 
for the anomalous dimension $\eta$ as a function 
of $G=\frac{\alpha_{s}}{8\pi}$ for $\Nc=3$ and $\Nf=3$ in $d=4$ dimensions.
For pure gluodynamics (i.e. $\Nf=0$), we find an 
IR stable fixed point for vanishing temperature,
\begin{equation}
\alpha_{*}=[\alpha_{*,8},\alpha_{*,3}]\approx[5.7,9.7],
\end{equation}
in agreement with the results found in \cite{Gies:2002af}.  The
(theoretical) uncertainty is due to the fact that we have used
a simple approximation for the exact color factors
$\tau_{j}^{A}$ and $\tau_{j}^{\psi}$, see App.  \ref{sec:color} for
details. This approximation introduces an artificial dependence on
the color direction of the background field. The extremal cases of
this dependence are given by the 3- and 8-direction in the Cartan
sub-algebra, the results of which span the above interval for the
IR fixed point. Even though this uncertainty is quantitatively large
in the pure-glue case, it has little effect on the quantitative
results for full QCD, see below.

The inclusion of light quarks yields a lower value for the
infrared fixed point $\alpha_{*}$, as can be seen from Fig.
\ref{fig:etaT0}. However, this lower fixed point will only be
attained if quarks stay massless or light in the deep IR. If \xsb\ 
occurs, the quarks become massive and decouple from the flow, such
that the system is expected to approach the pure-glue fixed point.
In any case, we can read off from Fig.  \ref{fig:etaT0} that,
already in the symmetric regime, the inclusion of quarks leads to a
smaller coupling $\alpha _s$ for scales $k>k_{\chi SB}$, as compared
to the coupling of a pure gluonic system.

\subsection{\label{sec:coup_res}Running-coupling results}

For quantitative results on the running coupling, we confine ourselves
to $d=4$ dimensions and to the gauge groups SU(2) and SU(3). Of
course, results for arbitrary $d>2$ and other gauge groups can
straightforwardly be obtained from our general expressions in
App. \ref{sec:Resummation}.\footnote{For instance, this offers a way
to study nonperturbative renormalizability of QCD-like theories in extra dimensions as
initiated in \cite{Gies:2003ic} for pure gauge theories.}  

\begin{figure}[t]
\includegraphics[%
  clip,
  scale=1.05]{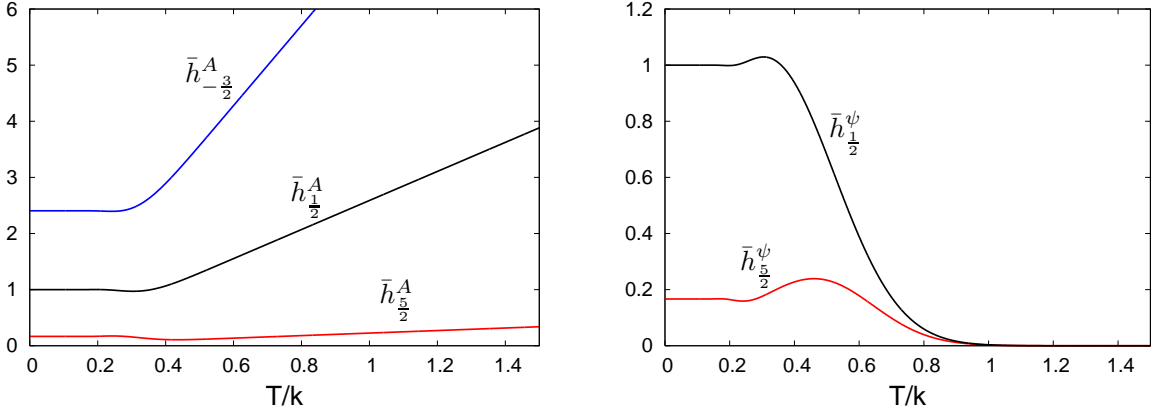}
\caption{Thermal moments as a function of $\frac{T}{k}$ for the 
exponential regulator.
The moments $h_{i}^{A}$ as well as $h_{i}^{\psi}$ are finite in the
limit $\frac{T}{k}\rightarrow 0$.  The gluonic thermal moments
$h_{i}^{A}$ grow linearly for increasing $\frac{T}{k}$ due to the
presence of a soft Matsubara mode, whereas the fermionic thermal
moments $h_{i}^{\psi}$ are exponentially supressed for
$\frac{T}{k}\rightarrow \infty$.}
\label{fig:moments} 
\end{figure}

To this end, a quantitative evaluation of the coupling flow requires
the specification of the regulator shape function $r(y)$,
cf.~Eq.~\eqref{eq:reg_def}. In order to make simple contact with
measured values of the coupling, e.g., at the $Z$ mass or the $\tau$
mass, it is advantageous to choose $r(y)$ in correspondence with a
regularization scheme for which the running of the coupling is
sufficiently close to the standard $\overline{\text{MS}}$ running in
the perturbative domain. Here, it is important to note that already
the two-loop $\beta_{g^2}$ coefficient depends on the regulator, owing
to both the truncation as well as the mass-dependent regularization
scheme. As an example, we give the two-loop $\beta_{g^2}$ function
calculated from Eq.~\eqref{eq:fe_wpot} for QCD with $\Nc$ colors and
$\Nf$ massless quark flavors in $d=4$ dimensions:
\begin{eqnarray}
\beta(g^{2})&=&-{\Bigg(\frac{22}{3}\bar{h}_{\frac{1}{2}}^{A}\Nc
-\frac{4}{3}\bar{h}_{\frac{1}{2}}^{\psi}\Nf\Bigg)
\frac{g^{4}}{(4\pi)^{2}}}- \Bigg(\,\frac{77\Nc^{2}
\bar{h}_{\frac{1}{2}}^{A} -
14\Nc\Nf\bar{h}_{\frac{1}{2}}^{\psi}}{3}
\,\bar{g}_{\frac{1}{2}}^{A}\\ 
& &\hspace*{-0.7cm}\qquad\quad
- \frac{127\tau _{2}^{A}\bar{h}_{\frac{5}{2}}^{A}
 + \Nf\tau _{2}^{\psi}\bar{h}_{\frac{5}{2}}^{\psi}}{45}
\Big(3(\Nc^{2}-1)(\bar{h}_{-\frac{3}{2}}^{A}
-\bar{g}_{-\frac{3}{2}}^{A})+2(\bar{H}_{0}^{A}-\bar{G}_{0}^{A})\Big)\Bigg)
\frac{g^{6}}{(4\pi)^{4}}+\nn \dots
\end{eqnarray}
The moments $\bar{g}_{j}^{A/\psi}$,$\bar{h}_{j}^{A/\psi}$,
$\bar{G}_{j}^{A}$ and $\bar{H}_{j}^{A}$ are defined in App.
\ref{sec:Regulator_function}. They specify the regulator dependence of
the loop terms and depend on $\frac{T}{k}$, as is visualized in
Fig.~\ref{fig:moments}.
We observe that even the one-loop coefficient
is regulator dependent at finite temperature, but universal and exact
at zero temperature, as it should. The latter holds, since
$\bar{g}_{\frac{1}{2}}^{A/\psi}(\frac{T}{k}=0)=1$ and
$\bar{h}_{\frac{1}{2}}^{A/\psi}(\frac{T}{k}=0)=1$ for all admissible
regulators. Using the exponential regulator, we find
$\bar{h}_{-\frac{3}{2}}^{A}(\frac{T}{k}=0)=2\zeta(3)$,
$\bar{g}_{-\frac{3}{2}}^{A}(\frac{T}{k}=0)=1$,
$\bar{h}_{\frac{5}{2}}^{A/\psi}(\frac{T}{k}=0)=\frac{1}{6}$,
$\bar{G}_{0}^{A}(\frac{T}{k}=0)=\frac{1}{2}$ and
$\bar{H}_{0}^{A}(\frac{T}{k}=0)=\zeta(3)$ for the moments at zero
temperature. Using the color factors $\tau _{2}^{A}$ and $\tau
_{2}^{\psi}$ from App.~\ref{sec:color}, we compare our result to the
perturbative two-loop result,
\begin{equation}
\beta_{pert.}(g^{2})=-{\Bigg(\frac{22}{3}\Nc-\frac{4}{3}\Nf\Bigg)\frac{g^{4}}{(4\pi)^{2}}} - \Bigg(\frac{68 \Nc^{3} + 6 \Nf - 26 \Nc^{2}\Nf}{3 \Nc}\Bigg)\frac{g^{6}}{(4\pi)^{4}}+\dots\,,
\end{equation}
and find good agreement to within 99\% for the two-loop coefficient
for SU(2) and 95\% for SU(3) pure gauge theory.  Beside this
compatibility with the standard $\overline{\mathrm{MS}}$ running the
exponential regulator is technically and numerically convenient.

The perturbative quality of the regulator is mandatory for
a reliable estimate of absolute, i.e., dimensionful, scales of the
final results. The present choice enables us to fix the running
coupling to experimental input: as initial condition, we use the
measured value of the coupling at the $\tau$ mass scale
\cite{Bethke:2004uy}, $\alpha_{\mathrm{s}}=0.322$, which by RG
evolution agrees with the world average of $\alpha_{\mathrm{s}}$ at
the $Z$ mass scale.  We stress that no other parameter or scale is
used as an input.

\begin{figure}[t]
\begin{center}
\includegraphics[%
  clip,
  scale=0.9]{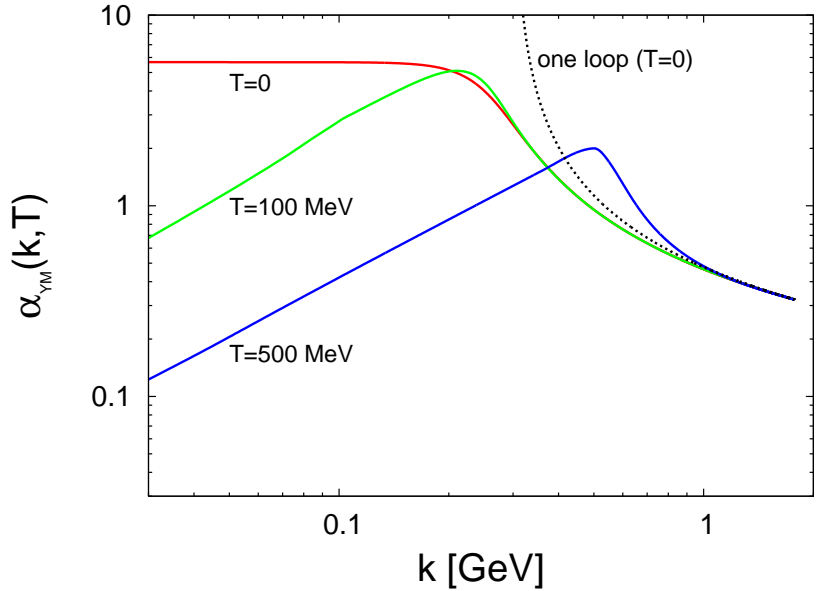}
\end{center}
\caption{Running $\mathrm{SU}(3)$ Yang-Mills coupling $\alpha
_{\scriptscriptstyle{\mathrm{YM}}} (k,T)$ as a function of $k$ for
$T=0,100,500\,\mathrm{MeV}$ compared to the one-loop running for
vanishing temperature.}
\label{fig:coupl_pure_glue} 
\end{figure}

The global behavior of the running coupling can be characterized in
simple terms. Let us first concentrate on pure gluodynamics, setting
$\Nf\to0$ for a moment. At zero temperature, we rediscover the results
of \cite{Gies:2002af}, exhibiting a standard perturbative behavior in
the UV. In the IR, the coupling increases and approaches a stable
fixed point $g^2_\ast$ which is induced by a second zero of the
$\beta_{g^2}$ function, see Fig.~\ref{fig:coupl_pure_glue}.
The appearance of an IR fixed point in
Yang-Mills theories is a well-investigated phenomenon also in the
Landau gauge \cite{vonSmekal:1997is}.
Here, the IR fixed point is a consequence of a tight link between the
fully dressed gluon and ghost propagators at low momenta which is
visible in a vertex expansion \cite{Alkofer:2004it}. Most
interestingly, this behavior is in accordance with the
Kugo-Ojima and Gribov-Zwanziger confinement scenarios
\cite{Kugo:1979gm}. Even though the relation between the Landau-gauge
and the background-gauge IR fixed point is not immediate, it is
reassuring that the definition of the running coupling in both
frameworks rests on a nonrenormalization property that arises from
gauge invariance \cite{Taylor:1971ff,Abbott:1980hw}. Within the present
mass-dependent RG scheme, the appearance of an IR fixed point is
moreover compatible with the existence of a mass gap: once the scale
$k$ has dropped below the lowest physical state in the spectrum, the
running of physically relevant couplings should freeze out, since no
fluctuations are left to drive any further RG flow. Finally, IR
fixed-point scenarios have successfully been applied also in
phenomenological studies
\cite{Dokshitzer:1998pt,Eichten:1974af,Grunberg:2000ap,%
Shirkov:1997wi,Brodsky:2002nb,Deur:2005cf}. 

At finite temperature, the small-coupling UV behavior remains
unaffected for scales $k\gg T$ and agrees with the zero-temperature
perturbative running as expected. Towards lower scales, the coupling
increases until it develops a maximum near $k\sim T$. Below, the
coupling decreases according to a powerlaw $g^2 \sim k/T$, see
Fig.~\ref{fig:coupl_pure_glue}. This behavior has a simple explanation:
the wavelength of fluctuations with momenta $p^2<T^2$ is larger than
the extent of the compactified Euclidean time direction. Hence, these
modes become effectively 3-dimensional and their limiting behavior is
governed by the spatial 3$d$ Yang-Mills theory. As a nontrivial 
result, we observe the existence of a non-Gau\ss ian IR fixed point
also in the reduced 3$d$ theory, see also Sec. \ref{sec:dim_red}. By virtue of a
straightforward matching between the 4$d$ and 3$d$ coupling, the
observed powerlaw for the 4$d$ coupling is a direct consequence of the
strong-coupling 3$d$ IR behavior, $g^2(k\ll T)\sim g^2_{3d,\ast}\,
k/T$. Again, the observation of an IR fixed point in
the 3$d$ theory agrees with recent results in the Landau gauge
\cite{Maas:2005hs}.
The 3$d$ IR fixed point and the perturbative UV behavior already
qualitatively determine the momentum asymptotics of the running
coupling. Phenomenologically, the behavior of the coupling in the
transition region near its maximum value is most important, which is
quantitatively provided by the full 4$d$ finite-temperature flow
equation. In addition to the shift of the position of the maximum with
temperature, we observe a decrease of the maximum itself for
increasing temperature.  On average, the 4$d$ coupling gets weaker for
higher temperature, in agreement with naive expectations. We
emphasize, however, that this behavior results from a nontrivial
interplay of various nonperturbative contributions.\medskip
\begin{figure}[t]
\begin{center}
\includegraphics[%
  clip,
  scale=0.9]{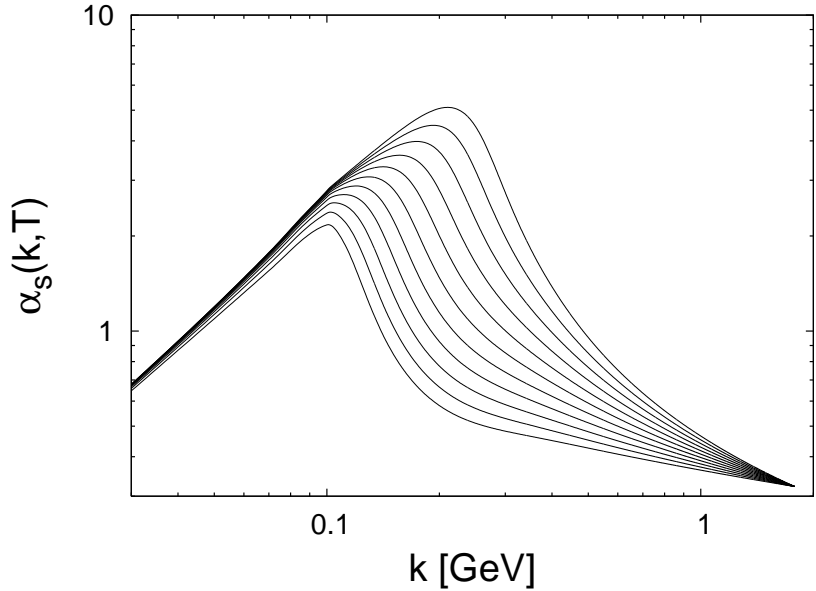}
\end{center}
\caption{Running $\mathrm{SU}(3)$ coupling $\alpha _{s} (k,T)$ as a
function of $k$ for $T=100\,\mathrm{MeV}$ for different number of
quark flavors $\Nf=0,1,2,\dots,10$ (from top to bottom). For $k\ll T$,
the coupling shows universal behavior, owing to the attraction of the
pure-glue IR fixed point.}
\label{fig:coupl_nf_t100} 
\end{figure}

Now, we turn to the effect of a finite number $\Nf$ of massless quark
flavors. In Fig. \ref{fig:coupl_nf_t100}, we show the running coupling
$\alpha_{s}$ as a function of $k$ for $T=100~\mathrm{MeV}$ and for
$\Nf=0, \dots, 10$.  At high scales $k\gg T$, the running of the
coupling agrees with the zero-temperature running in the presence of
$\Nf$ massless quark flavors. Towards lower scales, the coupling
increases less strongly than the coupling of the corresponding SU(3)
Yang-Mills theory, due to fermionic screening.  At a scale $k\sim T$,
the coupling reaches its maximum. Below this scale, the quarks
decouple from the flow, since they only have hard Matsubara modes and,
hence, the coupling universally approaches the result for pure
Yang-Mills theory.  Furthermore, we observe that, for an increasing
number of quark flavors, the maximum of the coupling becomes smaller
and moves towards lower scales. Both effects are due to the fact that
the anomalous dimension $\eta$ becomes smaller for an increasing
number of quark flavors.

Again, we stress that the results for the coupling with dynamical
quarks have not yet accounted for \xsb, where the quarks become
massive and decouple from the flow.  This will be discussed in the
following sections. For temperatures or flavor numbers larger than the
corresponding critical value for \xsb, our results so far should be
trustworthy on all scales.

\subsection{\label{sec:dim_red}Dimensionally reduced high-temperature
  limit} 

As discussed above, the running coupling for scales much lower than
the temperature, $k\ll T$, is governed by the IR fixed point of the
3-dimensional theory. More quantitatively, we observe that the
flow of the coupling is completely determined by $\eta_{1}^{A}$ for
$\frac{T}{k}\gg1$; the quark contributions decouple from the flow in
this limit, since they do not have a soft Matsubara mode. Therefore, we
find an IR fixed point at finite temperature for the 4$d$ theory at
$g^2=0$. In the limit $\frac{T}{k}\gg 1$, the anomalous dimension 
Eq.~\eqref{eq:etares} is given by
\begin{equation}
\eta (T\gg k)\approx\eta_{1}^{A}(T\gg
  k)=:\eta_{1}^{\infty}(g^2,\textstyle{\frac{T}{k}}) 
=\bar{\gamma}_{3d}\,\left(\textstyle{\frac{T}{k}}\,g^{2}\right)^{\frac{5}{4}},
\end{equation}
where $\bar{\gamma} _{3d}$ is a number which depends on $\Nc$:
\begin{equation}
\bar{\gamma}_{3d}
=\frac{32\zeta(\frac{5}{2})(1 -
  2\sqrt{2})\Gamma(\frac{9}{4})\Gamma(\frac{5}{4}+z_{4}^{\infty})
\sqrt[4]{{c}_{1}^{\infty}}}{(4\pi)^{4}\Gamma(\frac{3}{2})\Gamma(z_{4}^{\infty}+1)}\Nc. 
\end{equation}
We refer to App. \ref{sec:Resummation} for the definition of the constants $z_4 ^{\infty}$ and ${c}_1^{\infty}$.
In the high-temperature limit, we can solve the differential equation \eqref{betadef} 
for $g^{2}$ analytically,
\begin{equation}
g^{2}\Big|_{\frac{T}{k}\gg1}=:g_{\infty}^{2}({\textstyle{\frac{k}{T}}})
=\frac{1}{(\bar{\gamma}_{3d}(\frac{T}{k})^{\frac{5}{4}}-\mathrm{{const.}})^{\frac{4}{5}}}\,
\approx\,\bar{\gamma}_{3d} ^{-\frac{4}{5}}\,{\frac{k}{T}}
+\mathcal{O}((\textstyle{\frac{k}{T}})^2).
\label{eq:g2infty}
\end{equation}
The RHS explains the shape of the running coupling for small $k/T$ in
Fig. \ref{fig:coupl_pure_glue}.  The factor
$\bar{\gamma}_{3d}^{-\frac{4}{5}}$ is the fixed point value of the
dimensionless 3$d$ coupling ${g}_{3d}^{2}$, as can be seen from
its relation to the dimensionless coupling $g^2$ in four dimensions:
\begin{equation}
g_{3d}^{2}:=\frac{T}{k}\, g^{2} \quad \rightarrow g^{2}
=\frac{k}{T}\,{g}_{3d}^{2}\,.\label{eq:g4d3d}
\end{equation}
Comparing the right-hand side of Eq.~\eqref{eq:g2infty} and
\eqref{eq:g4d3d}, we find that the fixed point for $\Nc=3$ in three
dimensions is given by:
\begin{equation}
\alpha_{*}^{3d}\equiv\frac{{g}_{3d,*}^{2}}{4\pi}
=[\alpha_{*,8}^{3d},\alpha_{*,3}^{3d}]\approx[2.70,2.77]
\end{equation}
Again, the uncertainty arises from our ignorance of the exact
color factors $\tau_{m}^{A}$, see App. \ref{sec:Resummation} and App.
\ref{sec:color}.

On the other hand, the fixed point of the 3d theory
is determined by the zero of the corresponding $\beta$ function.
In fact, $\eta_{1}^{\infty}(g^2,\textstyle{\frac{T}{k}})$ is
identical to the 3d anomalous dimension $\eta_{3d}({g}_{3d}^{2})$, 
as can be deduced from the pure $3d$ theory, and we obtain
\begin{equation}
\pat(\textstyle{\frac{T}{k}}g^2)\equiv\pat{g}_{3d}^{2}
=(\eta_{3d}({g}_{3d}^{2})-1){g}_{3d}^{2}
, 
\end{equation}
as suggested by Eq. \eqref{betadef}. Since $\eta_{3d}$ is a
monotonously increasing function, we find a $3d$ IR fixed point for
${g}_{3d,*}^{2}=\bar{\gamma}_{3d}^{-\frac{4}{5}}$ which
coincides with the result above.

\section{Chiral quark dynamics}
\label{sec:quarks}

Dynamical quarks influence the QCD flow by two qualitatively
different mechanisms. First, quark fluctuations directly
modify the running coupling as already discussed above; the
nonperturbative contribution in the form of $\eta^{\text{q}}$ in
Eq.~\eqref{eq:etares} accounts for the screening nature of fermionic
fluctuations, generalizing the tendency that is already visible in
perturbation theory. Second, gluon exchange between quarks induces
quark self-interactions which can become relevant in the strongly
coupled IR. Both the quark and the gluon sector feed back onto each
other in an involved nonlinear fashion. In general, these
nonlinearities have to be taken into account and are provided by the
flow equation. However, we will argue that some intricate
nonlinearities drop out or are negligible for locating the chiral
phase boundary in a first approximation. 

Working solely in $d=4$ from here on, let us now specify the last 
part of our truncation: the effective action of quark
self-interactions $\Gamma_k^{\text{q-int}}[\yb,\psi]$, introduced in
Eq.~\eqref{eq:quarktrunc1}. In a consistent and systematic operator
expansion, the lowest nontrivial order is given by \cite{Gies:2003dp}
\begin{eqnarray}
\Gamma_k&=&\int_x \frac{1}{2} \Big[
  \bar\lambda_-(\text{V--A}) +\bar\lambda_+ (\text{V+A})
  +\bar\lambda_\sigma (\text{S--P}) +\bar\lambda_{\text{VA}}
  [2(\text{V--A})^{\text{adj}}\!+({1}/{\Nc})(\text{V--A})] \Big].
\label{equ::truncsym}
\end{eqnarray}
The four-fermion interactions occurring here have been classified
according to their color and flavor structure. Color and flavor
singlets are
\begin{eqnarray}
(\text{V--A})&=&(\yb\gamma_\mu\psi)^2 + (\yb\gamma_\mu\gamma_5\psi)^2,
\\
(\text{V+A}) &=&(\yb\gamma_\mu\psi)^2 - (\yb\gamma_\mu\gamma_5\psi)^2 ,
\end{eqnarray}
where (fundamental) color ($i,j,\dots$) and flavor ($\chi,\xi,\dots$)
indices are contracted pairwise, e.g., $(\yb\psi)\equiv (\yb_i^{\chi}
\psi_i^{\chi})$.  The remaining operators have non-singlet color or flavor
structure,
\begin{eqnarray}
(\text{S--P})&=&(\yb^{\chi}\psi^{\xi})^2-(\yb^{\chi}\gamma_5\psi^{\xi})^2
\equiv
   (\yb_i^{\chi}\psi_i^{\xi})^2-(\yb_i^{\chi}\gamma_5\psi_i^{\xi})^2,\nonumber\\
(\text{V--A})^{\text{adj}}&=&(\yb \gamma_\mu T^a\psi)^2 
   + (\yb\gamma_\mu\gamma_5 T^a\psi)^2, \label{eq::colorflavor}
\end{eqnarray}
where $(\yb^{\chi}\psi^{\xi})^2\equiv \yb^{\chi}\psi^{\xi} 
\yb^{\xi} \psi^{\chi}$, etc., and
$(T^a)_{ij}$ denotes the generators of the gauge group in the
fundamental representation.  The set of fermionic self-interactions
introduced in Eq.~\eqref{equ::truncsym} forms a complete basis. Any
other pointlike four-fermion interaction which is invariant under
\mbox{$\textrm{SU}(\Nc)$} gauge symmetry and
$\textrm{SU}(\Nf)_{\textrm{L}}\times \textrm{SU}(\Nf)_{\textrm{R}}$
flavor symmetry is reducible by means of Fierz transformations.
U${}_{\text{A}}(1)$-violating interactions are neglected, since we
expect them to become relevant only inside the \xsb\ regime or for
small $\Nf$; since the lowest-order U${}_{\text{A}}$(1)-violating term
schematically is $\sim (\yb\psi)^{\Nf}$, larger $\Nf$ correspond to
larger RG ``irrelevance'' by naive power-counting. For $\Nf=1$,
such a term is, of course, important, since it provides for a direct
fermion mass term; in this case, the chiral transition is expected to be a
crossover. Dropping the U${}_{\text{A}}(1)$-violating interactions, we
thus confine ourselves to $\Nf\geq2$.

We emphasize that the $\bar\lambda$'s are not considered as
independent external parameters as, e.g., in the Nambu--Jona-Lasinio
model. More precisely, we impose the boundary condition
$\bar\lambda_i\to 0$ for $k\to\Lambda\to\infty$ which guarantees that the
$\bar\lambda$'s at $k<\Lambda$ are solely generated by quark-gluon
dynamics, e.g., by 1PI ``box'' diagrams with 2-gluon exchange.
 
As a severe approximation, we drop any nontrivial momentum
dependencies of the $\bar\lambda$'s and study these couplings in the
point-like limit $\bar\lambda(|p_i|\ll k)$. This inhibits a study of
QCD properties in the chirally broken regime, since mesons, for
instance, manifest themselves as momentum singularities in the
$\bar\lambda$'s.  Nevertheless, the point-like truncation can be a
reasonable approximation in the chirally symmetric regime; this has
recently been quantitatively confirmed for the zero-temperature chiral
phase transition in many-flavor QCD \cite{Gies:2005as}, where the
regulator independence of universal quantities has been shown to hold
remarkably well even in this restrictive truncation. By adopting the
same system at finite $T$, we base our truncation on the assumption
that quark dynamics both near the finite-$T$ phase boundary as well as
near the many-flavor phase boundary \cite{Banks:1981nn} is driven by
qualitatively similar mechanisms.

The resulting flow equations for the $\bar\lambda$'s are a
straightforward generalization of those derived and analyzed in
\cite{Gies:2003dp,Gies:2005as} to the case of finite temperature.
Introducing the dimensionless renormalized couplings
\begin{equation}
\lambda_i={k^{2}\bar{\lambda_i}},
\end{equation}
(recall that $Z_\psi=1$ in our truncation), the flows of the quark
interactions read
\begin{eqnarray}
\pat\lm
&=& 2\lm\!
    -4v_4 \lFBo\left[ \frac{3}{\Nc}g^2\lm
            -3g^2 \lva \right]
-\frac{1}{8}v_{4}\lFB\left[\frac{12+9\Nc^2}{\Nc^2}g^{4} \right]
    \label{eq:lm}\\
&&\!\!-8 v_4\lF \Big\{-\Nf\Nc(\lm^2+\lp^2) + \lm^2
-2(\Nc+\Nf)\lm\lva   +\Nf\lp\lsf + 2\lva^2 \Big\},
\nonumber\\
\!\pat\lp&=&2 \lp\! -4v_4\lFBo \left[
-\frac{3}{\Nc}g^2\lp\right]
  -\frac{1}{8}v_{4}\lFB\left[
    -\frac{12+3\Nc^2}{\Nc^2} g^{4} \right]  \label{eq:lp}
\\
&&-8 v_4 \lF \Big\{ - 3\lp^2 - 2\Nc\Nf\lm\lp
- 2\lp(\lm+(\Nc+\Nf)\lva)
        + \Nf\lm\lsf
\nonumber\\
&&\!\!\quad\quad\quad\qquad+ \lva\lsf
        +\case{1}{4}\lsf{}^2 \Big\},
    \nonumber\\
\pat\lsf
&=&
    2\lsf\! -4v_4 \lFBo \left[6\Cas\, g^2\lsf
    -6g^2\lp \right]
    -\frac{1}{4} v_4 \lFB \Big[ -\frac{24
    -9\Nc^2}{\Nc}\, g^4  \Big]\label{eq:lsf}\\
&& -8 v_4 \lF
      \Big\{ 2\Nc \lsf^2\!  -\! 2\lm\lsf\! - 2\Nf\lsf\lva\!\!
- \!6\lp\lsf\! \Big\},
        \nonumber\\
\pat\lva
&=& 2 \lva\!-4v_4 \lFBo \left[
\frac{3}{\Nc}g^2\lva -3g^2\lm \right]
    -\frac{1}{8} v_4 \lFB \left[ -\frac{24 - 3\Nc^2}{\Nc}
g^4 \right]\label{eq:lva} \\
&& -8 v_4 \lF\!
  \Big\{\! - \!(\Nc+\Nf)\lva^2\! + 4\lm\lva\!
        - \case{1}{4} \Nf \lsf^2\Big\}.\nonumber
\end{eqnarray}
Here, $\Cas=(\Nc^2-1)/(2\Nc)$ is a Casimir operator of the gauge group,
and $v_4=1/(32\pi^2)$. For better readability, we have written all
gauge-coupling-dependent terms in square brackets, whereas fermionic
self-interactions are grouped inside braces. The threshold functions
$\lF,\lFB,\lFBo$ depend on the details of the regularization, see App.
\ref{sec:Regulator_function}; for zero quark mass and vanishing
temperature, these functions reduce to simple positive numbers,
see, e.g., Eqs.~\eqref{eq:lF1} and \eqref{eq:lFB}.\footnote{Here,
we ignore a weak dependence of the threshold functions on the anomalous
quark and gluon dimensions which were shown to influence the
quantitative results for the present system only on the percent
level, if at all \cite{Gies:2005as}.} For quark masses and
temperature becoming larger than the regulator scale $k$, these
functions approach zero, which reflects the decoupling of massive modes
from the flow.

Within this set of degrees of freedom, a simple picture for the chiral
dynamics arises: for vanishing gauge coupling, the flow is solved by
vanishing $\lambda_i$'s, which defines the Gau\ss ian fixed point. This
fixed point is IR attractive, implying that these self-interactions
are RG irrelevant for sufficiently small bare couplings, as they
should. At weak gauge coupling, the RG flow generates quark
self-interactions of order $\lambda\sim g^4$, as expected for a
perturbative 1PI scattering amplitude. The back-reaction of these
self-interactions on the total RG flow is negligible at weak
coupling. If the gauge coupling in the IR remains smaller than a
critical value $g<g_{\text{cr}}$, the self-interactions remain
bounded, approaching fixed points in the IR. These fixed points can
simply be viewed as order-$g^4$ shifted versions of the Gau\ss ian
fixed point, being modified by the gauge dynamics. At these fixed
points, the fermionic subsystem remains in the chirally invariant
phase which is indeed realized at high temperature.

\begin{figure}[!t]
\begin{center}
\scalebox{0.75}[0.75]{
\begin{picture}(190,160)(40,0)
\includegraphics[width=12.5cm,height=7cm]{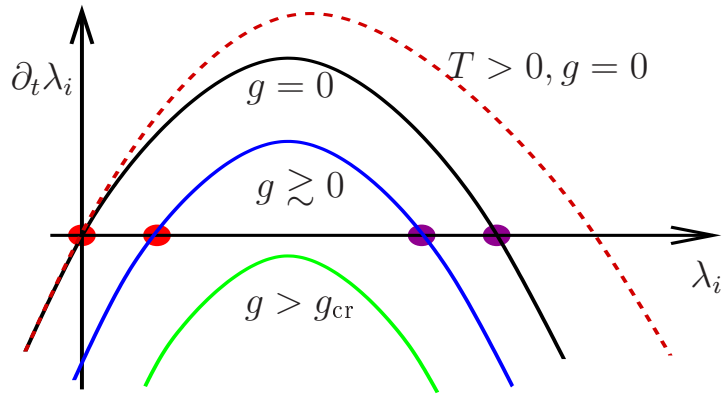}
\Text(-10,+60)[c]{\scalebox{1.6}[1.6]{$\lambda_i$}}
\Text(-345,160)[c]{\scalebox{1.6}[1.6]{$\pat\lambda_i$}}
\Text(-220,155)[c]{\scalebox{1.6}[1.6]{$g=0$}}
\Text(-215,105)[c]{\scalebox{1.6}[1.6]{$g\gtrsim 0$}} 
\Text(-215,45)[c]{\scalebox{1.6}[1.6]{$g>g_{\text{cr}}$}}
\Text(-90,165)[c]{\scalebox{1.6}[1.6]{$T>0,g=0$}} 
\end{picture}
}
\end{center}
\caption{Sketch of a typical $\beta$ function for the fermionic
  self-interactions $\lambda_i$: at zero gauge coupling, $g=0$ (upper
  solid curve), the Gau\ss ian fixed point $\lambda_i=0$ is IR
  attractive. For small $g\gtrsim 0$ (middle/blue solid curve), the
  fixed-point positions are shifted on the order of $g^4$. For gauge
  couplings larger than the critical coupling $g>g_{\text{cr}}$
  (lower/green solid curve), no fixed points remain and the
  self-interactions quickly grow large, signaling \xsb. For increasing
  temperature, the parabolas become broader and higher, owing to
  thermal fermion masses; this is indicated by the dashed/red 
  line.} 
\label{fig:parab}
\end{figure}

If the gauge coupling increases beyond the critical coupling
$g>g_{\text{cr}}$, the above-mentioned IR fixed points are
destabilized and the quark self-interactions become critical. This can
be visualized by the fact that $\pat \lambda_i$ as a function of
$\lambda_i$ is an everted parabola, see Fig. \ref{fig:parab}; for
$g=g_{\text{cr}}$, the parabola is pushed below the $\lambda_i$ axis,
such that the (shifted) Gau\ss ian fixed point annihilates with the
second zero of the parabola. In this case, the
gauge-fluctuation-induced {$\bar\lambda$}'s have become strong enough to
contribute as relevant operators to the RG flow. These couplings now
increase rapidly, approaching a divergence at a finite scale
$k=k_{\text{\xsb}}$. In fact, this seeming Landau-pole behavior
indicates \xsb\ and, more specifically, the formation of chiral
condensates. This is because the {$\bar\lambda$}'s are proportional to
the inverse mass parameter of a Ginzburg-Landau effective potential
for the order parameter in a (partially) bosonized formulation,
{$\bar\lambda\sim 1/m^2$}. Thus, the scale at which the
self-interactions formally diverge in our truncation is a good measure
for the scale $k_{\text{\xsb}}$ where the effective potential for the
chiral order parameter becomes flat and is about to develop a nonzero
vacuum expectation value.

Whether or not chiral symmetry is preserved by the ground state
therefore depends on the coupling strength of the system, more
specifically, the value of the gauge coupling $g$ relative to the
critical coupling $g_{\text{cr}}$ which is required to trigger
\xsb. Incidentally, the critical coupling $g_{\text{cr}}$ itself can
be determined by algebraically solving the fixed-point equations $\pat
\lambda_i(\lambda_\ast)=0$ for that value of the coupling,
$g=g_{\text{cr}}$, where the shifted Gau\ss ian fixed point is
annihilated. For instance, at zero temperature, the SU(3) critical
coupling for the quarks system is $\alpha_{\text{cr}}\equiv
g^2_{\text{cr}}/(4\pi)\simeq 0.8$ \cite{Gies:2002hq}, being only
weakly dependent on the number of flavors
\cite{Gies:2005as}.\footnote{Of course, the critical coupling is a
non-universal value depending on the regularization scheme; the value
given here for illustration holds for a class of regulators in the
functional RG scheme that includes the most widely used linear
(``optimized'') and exponential regulators.}  Since the IR fixed point
for the gauge coupling is much larger $\alpha_\ast>\alpha_{\text{cr}}$
(for not too many massless flavors), the QCD vacuum is characterized
by \xsb. The same qualitative observations have already been made in
\cite{Aoki:1999dw} in a similar though smaller truncation. The
existence of such a critical coupling also is a well-studied
phenomenon in Dyson-Schwinger equations \cite{Miransky:1984ef}.

As soon as the the quark sector approaches criticality, also its
back-reaction onto the gluon sector becomes sizable.
Here, a subtlety of the present formalism becomes important:
identifying the fluctuation field with the background field under the
flow, our approximation generally does not distinguish between the
flow of the background-field coupling and that of the
fluctuation-field coupling. In our truncation, differences arise
from the quark self-interactions. Whereas the running of the
background-field coupling is always given by Eq.~\eqref{betadef}, 
the quark self-interactions can contribute directly to the running of
the fluctuation-field coupling in the form of a ``vertex correction''
to the quark-gluon vertex. Since the fluctuation-field coupling
is responsible for inducing quark self-interactions, this difference
may become important. In \cite{Gies:2003dp}, the relevant terms
have been derived with the aid of a regulator-dependent Ward-Takahashi
identity. The result hence implements an important gauge constraint,
leading us to
\begin{eqnarray}
\partial_{t}g^2&=&\eta\, g^2-4 v_4
  \lF\, \frac{g^2}{1-2v_4\lF \sum c_i \lambda_i} \, \pat \sum c_i
  \lambda_i, \label{betaeq} \\ 
&&
c_{\sigma}=1+\Nf,\,\,c_{+}=0,\,\,c_{-}=-2,\,\,c_{\textrm{VA}}=-2\Nf,
\nonumber
\end{eqnarray}
with $\eta$ provided by Eq.~\eqref{eq:etares} in our approximation. In
principle, the approach to \xsb\ can now be studied by solving the
coupled system of Eqs.~\eqref{betaeq}, and
\eqref{eq:lm}-\eqref{eq:lva}. However, a simpler and, for our
purposes, sufficient estimate is provided by the following argument:
if the system ends up in the chirally symmetric phase, the
$\lambda_i$'s always stay close to the shifted Gau\ss ian fixed point
discussed above; apart from a slight variation of this fixed-point
position with increasing $g^2$, the $\pat\lambda_i$ flow is small and
vanishes in the IR, $\pat \lambda_i\to 0$. Therefore, the additional
terms in Eq.~\eqref{betaeq} are negligible for all $k$ and drop out
in the IR.  As a result, the behavior of the running coupling
in the chirally symmetric phase is basically determined by $\eta$
alone, as discussed in the preceding section. In other words, the
difference between the fluctuation-field coupling and the
background-field coupling automatically switches off in the deep IR in
the symmetric phase in our truncation.

Therefore, if the coupling as predicted by $\beta_{g^2}\simeq \eta
g^2$ alone never increases beyond the critical value $g^2_{\text{cr}}$
for any $k$, the system is in the chirally symmetric phase. In this
case, it suffices to solve the $g^2$ flow and compare it with
$g^2_{\text{cr}}$ which can be deduced from a purely algebraic
solution of the fixed-point equations, $\pat
\lambda_i(\lambda_\ast)=0$. 

If the coupling as predicted by $\beta_{g^2}\simeq \eta g^2$ alone
approaches $g_{\text{cr}}$ for some finite scale $k_{\text{cr}}$, the
quark sector becomes critical and all couplings start to flow
rapidly. To the present level of accuracy, this serves as an
indication for \xsb. Of course, if the gauge coupling dropped quickly
for decreasing $k$, the quark sector could, in principle, become
subcritical again. However, this might happen only for a marginal
range of $g^2\simeq g^2_{\text{cr}}$ if at all. For even larger gauge
coupling, the flow towards \xsb\ is unavoidable. 

Inside the \xsb\ regime, also the induced quark masses back-react onto
the gluonic flow in the form of a decoupling of the quark
fluctuations, i.e., $\eta^{\text{q}}$ in Eq.~\eqref{eq:etares}
approaches zero. However, the present truncation does not allow to
explore the properties of the \xsb\ sector; for this, the introduction
of effective mesonic degrees of freedom along the lines of
\cite{Gies:2002hq,Gies:2001nw} is most useful and will be employed in
future work.

\section{Chiral phase transition}
\label{sec:xsb}

Let us now discuss our results for the chiral phase transition in the
framework presented so far. As elucidated in the previous section, the
breaking of chiral-symmetry is triggered if the gauge coupling $g^2$
increases beyond $g^2_{\text{cr}}$, signaling criticality of the quark
sector. We study the dependence of the chiral symmetry status on two
parameters: temperature $T$ and number of (massless) flavors $\Nf$. As
already discussed in Sect.~\ref{sec:RunCoup}, the increase of the
running coupling in the IR is weakened on average for both larger $T$ and 
larger $\Nf$. In addition, also $g_{\text{cr}}$ depends on $T$ and
$\Nf$, even though the $\Nf$ dependence is rather weak. 

The $T$ dependence of $g_{\text{cr}}$ has a physical interpretation:
at finite $T$, all quark modes acquire thermal masses which leads to a
quark decoupling for $k\lesssim T$. Hence, stronger interactions are
required to excite critical quark dynamics. Technically, this $T/k$
dependence is a direct consequence of the $T/k$ dependence of the
threshold functions $\lF,\lFB,\lFBo$ in Eqs.~
\eqref{eq:lm}~-~\eqref{eq:lva}, see App. \ref{sec:Regulator_function} for their definition. 
Since the threshold functions decrease
with increasing temperature, the $\lambda_i$ parabolas visualized in
Fig. \ref{fig:parab} become broader with a higher maximum; hence, the
annihilation of the Gau\ss ian fixed point by pushing the parabola
below the $\lambda_i$ axis requires a larger $g_{\text{cr}}$.
\begin{figure}[t]
\includegraphics[scale=0.6]{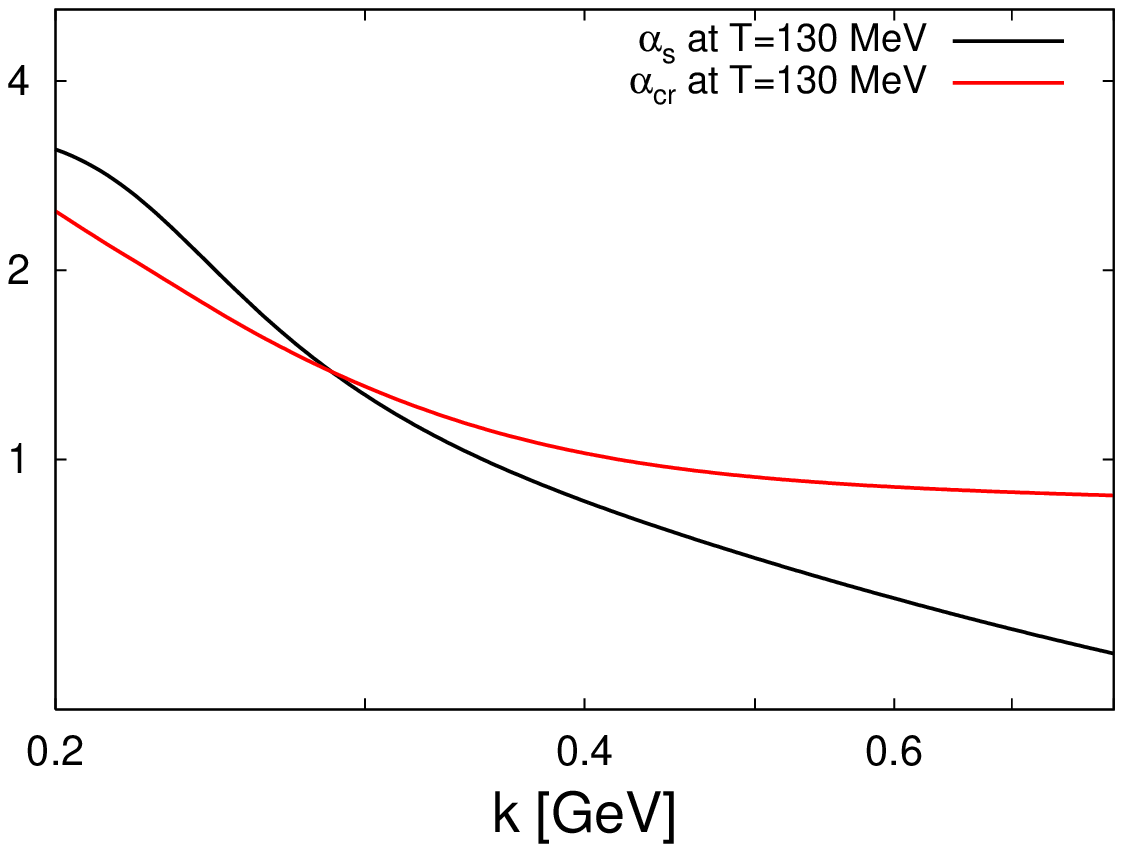}
\includegraphics[scale=0.6]{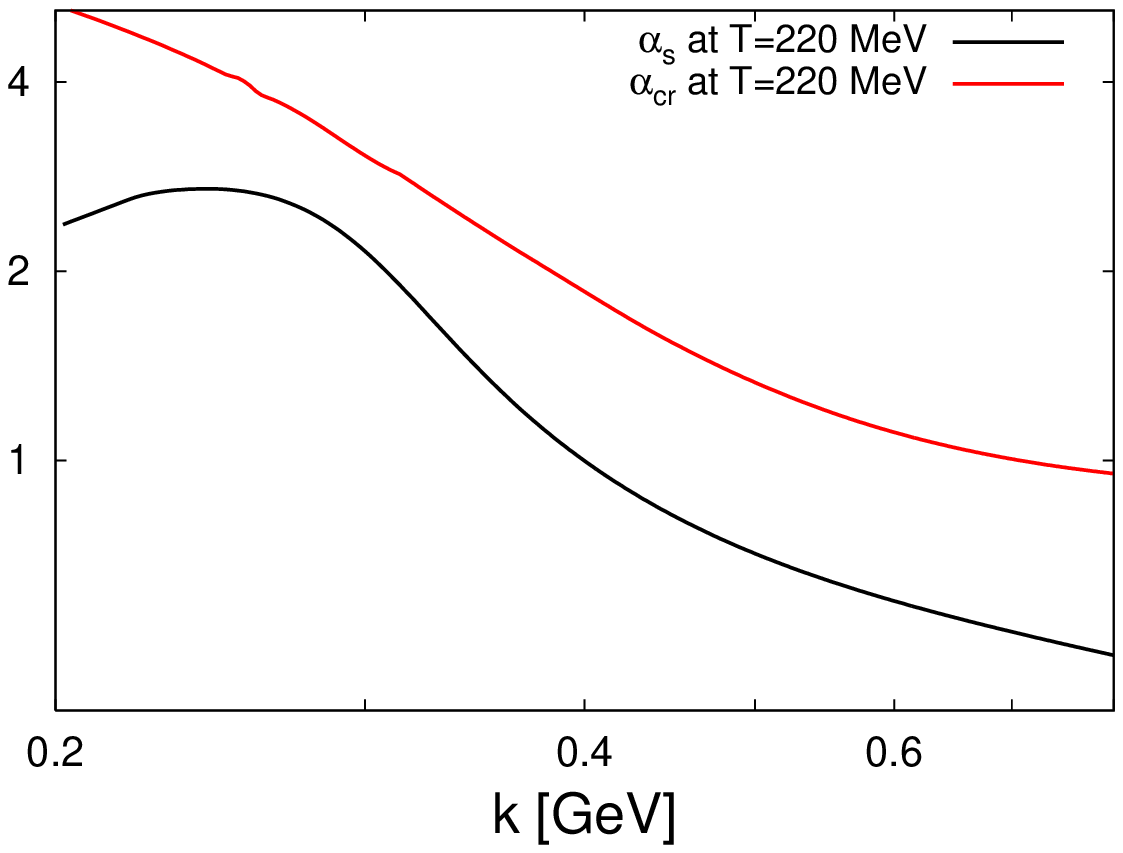}
\caption{Running QCD coupling
$\alpha_{\mathrm{s}} (k,T)$ for $\Nf=3$ massless quark flavors and
$\Nc=3$ colors and the critical value of the running coupling
$\alpha_{\mathrm{cr}} (k,T)$ as a function of $k$ for
$T=130\,\mathrm{MeV}$ (left panel) and $T=220\,\mathrm{MeV}$ (right
panel). The existence of the
$(\alpha_{\mathrm{s}},\alpha_{\mathrm{cr}})$ intersection point 
in the left panel indicates that the \xsb\ quark dynamics can 
become critical for $T=130\,\mathrm{MeV}$.}
\label{alpha_alpha_c} 
\end{figure}

\begin{figure}[t]
\begin{center}
\includegraphics[%
  clip,
  scale=0.9]{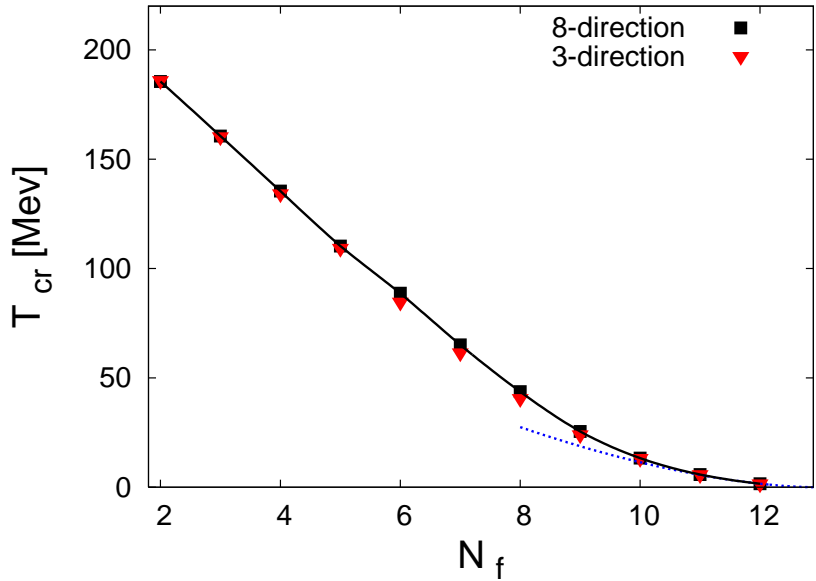}
\end{center}
\caption{Chiral-phase-transition temperature $T_{\text{cr}}$ versus
the number of massless quark flavors $\Nf$ for $\Nf\geq2$. The
flattening at $\Nf\gtrsim10$ is a consequence of the IR fixed-point
structure. The dotted line depicts the analytic estimate near
$\Nf^{\text{cr}}$ which follows from the fixed-point scenario
(cf. Eq.~\eqref{eq:TcrTheta} below). Squares and triangles correspond
to calculations with a background field in the 8- and 3-direction of
the Cartan, respectively. The theoretical uncertainty which is given
by the difference between both is obviously negligible in full QCD.}
\label{tc_nf} 
\end{figure}

At zero temperature and for small $\Nf$, the IR fixed point of the
running coupling is far larger than $g^2_{\text{cr}}$, hence the QCD
vacuum is in the \xsb\ phase. For increasing $T$, the temperature
dependence of the coupling and that of $g^2_{\text{cr}}$ compete with
each other. This is illustrated in Fig. \ref{alpha_alpha_c} where we
show the running coupling $\alpha_{\mathrm{s}}\equiv \frac{g^2}{4\pi}$ 
and its critical value $\alpha_{\mathrm{cr}}\equiv \frac{g_{cr} ^2}{4\pi}$ for $T=130\,\mathrm{MeV}$ and
$T=220\,\mathrm{MeV}$ as a function of the regulator scale $k$. The
intersection point $k_{\text{cr}}$ between both marks the scale where
the quark dynamics becomes critical. Below the scale $k_{\text{cr}}$,
the system runs quickly into the \xsb\ regime. We estimate the
critical temperature $T_{\text{cr}}$ as the lowest temperature for
which no intersection point between $\alpha_{\mathrm{s}}$ and
$\alpha_{\mathrm{cr}}$ occurs.\footnote{Strictly speaking, 
this simplified analysis yields a
sufficient but not a necessary criterion for chiral-symmetry
restoration. In this sense, our estimate for $T_{\text{cr}}$ is an
upper bound for the true $T_{\text{cr}}$. Small corrections to this
estimate could arise, if the quark dynamics becomes uncritical again
by a strong decrease of the gauge coupling towards the IR, as discussed
in the preceding section.}  We find
\begin{eqnarray}
T_{\mathrm{cr}}&\approx& 186\pm 33\,\mathrm{MeV\quad for\quad}\Nf=2,
  \nonumber\\ 
T_{\mathrm{cr}}&\approx& 161\pm 31\,\mathrm{MeV\quad for\quad}\Nf=3,
\label{eq:Tcres}
\end{eqnarray}
for massless quark flavors in good agreement with lattice simulations
\cite{Karsch:2000kv}.  The errors arise from the experimental
uncertainties on $\alpha_{\mathrm{s}}$
\cite{Bethke:2004uy}. The theoretical error owing to the color-factor 
uncertainty turns out to be subdominant by far, see Fig. \ref{tc_nf}.
Dimensionless observable ratios are less
contaminated by this uncertainty of $\alpha_{\text{s}}$. For instance,
the relative difference for $\Tc$ for $\Nf=2$ and 3 flavors is
\begin{equation}
\Delta:=\frac{\Tc^{\Nf=2}-\Tc^{\Nf=3}}{(\Tc^{\Nf=2}+\Tc^{\Nf=3})/2}
=0.144\genfrac{}{}{0pt}{}{+ 0.018}{- 0.013},\label{eq:Deltares}
\end{equation}
in reasonable agreement with the lattice value of $\sim
0.12$ \cite{Karsch:2000kv}.\footnote{Even this comparison is 
potentially contaminated by fixing the two theories with different flavor 
content in different ways. Whereas lattice simulations generically keep the
string tension fixed, we determine all scales by fixing $\alpha$
at the $\tau$ mass scale, cf. the discussion below.}

For the case of many massless quark flavors $\Nf$, the critical
temperature is plotted in Fig.~\ref{tc_nf}. We observe an almost
linear decrease of the critical temperature for increasing $\Nf$ with
a slope of $\Delta T_{\mathrm{cr}}=T(\Nf)-T(\Nf+1)\approx
25\,\mathrm{MeV}$.  In addition, we find a critical number of quark
flavors, $\Nf ^{\mathrm{cr}}\simeq12.9$, above which no chiral phase
transition occurs.  This result for $\Nf^{\mathrm{cr}}$ agrees with
other studies based on the 2-loop $\beta$ function
\cite{Banks:1981nn}. However, the precise value of $\Nf^{\mathrm{cr}}$
has to be taken with care: for instance, in a perturbative framework,
$\Nf^{\mathrm{cr}}$ is sensitive to the 3-loop coefficient which can
bring $\Nf^{\text{cr}}$ down to $\Nf^{\text{cr}}\simeq 10$
\cite{Gies:2005as}. In our nonperturbative approach, the truncation
error can induce similar uncertainties; in fact, it is reassuring that
our prediction for $\Nf^{\text{cr}}$ lies in the same ball park as the
perturbative estimates, even though the details of the corresponding
$\beta_{g^2}$ are very different. This suggests that our truncation
error for $\Nf^{\text{cr}}$ is also of order $\mathcal O (1)$. We
expect that a more reliable estimate can be obtained even within our
truncation by a regulator \cite{Litim:2001up,Pawlowski:2005xe}. 

A remarkable feature of the $T,\Nf$ phase diagram of Fig. \ref{tc_nf}
is the shape of the phase boundary, in particular, the flattening
near $\Nf^{\text{cr}}$. In fact, this shape can be understood
analytically, revealing a direct connection between two universal
quantities: the phase boundary and the IR critical exponent of the
running coupling. 

Before we outline the argument in detail, let us start with an
important caveat: varying $\Nf$ unlike varying $T$ corresponds to an
unphysical deformation of a physical system. Whereas the deformation
itself is, of course, unambiguously defined, the comparison of the
physical theory with the deformed theory (or between two deformed
theories) is not unique. A meaningful comparison requires to identify
one parameter or one scale in both theories. In our case, we always
keep the running coupling at the $\tau$ mass scale fixed to
$\alpha(m_\tau)=0.322$. Obviously, the couplings in the two theories
are different on all other scales, as are generally all dimensionful
quantities such as $\Lambda_{\text{QCD}}$. There is, of course, no
generic choice for fixing the corresponding theories relative to each
other. Nevertheless, we believe that our choice is particularly
useful, since the $\tau$ mass scale is close to the transition between
perturbative and nonperturbative regimes. In this sense, a meaningful
comparison between the theories can be made in both regimes, without
being too much afflicted by the choice of the fixing condition.

Let us now study the shape of the phase boundary for small
$\Nf$. Once the coupling is fixed to $\alpha(m_\tau)=0.322$, no free
parameter is left. As a crude approximation, the mass scale of all
dimensionful IR observables such as the critical temperature
$T_{\text{cr}}$ is set by the scale $k_{\text{co}}$ where the running
gauge coupling undergoes the crossover from small to nonperturbatively
large couplings (for instance, one can define the crossover scale
$k_{\text{co}}$ from the inflection point of the running coupling in
Fig. \ref{fig:coupl_pure_glue}). As an even cruder estimate, let us
approximate $k_{\text{co}}$ by the position of the Landau pole of the
perturbative one-loop running coupling.\footnote{Actually, this is a
reasonable estimate, since the $\Nf$ dependence of $k_{\text{co}}$,
which is all that matters in the following, is close to the
  perturbative behavior.} The latter can be derived from the one-loop
relation
\begin{equation}
\frac{1}{\alpha(k)}=\frac{1}{\alpha(m_\tau)} +4\pi b_0 \ln
\frac{k}{m_\tau}, \quad b_0=\frac{1}{8\pi^2} \left( \frac{11}{3} \Nc
-\frac{2}{3} \Nf \right). 
\end{equation}
Defining $k_{\text{co}}$ by the Landau-pole scale,
$1/\alpha(k_{\text{co}}) =0$, and estimating the order of the critical
temperature by $T_{\text{cr}}\sim k_{\text{co}}$, we obtain
\begin{equation}
T_{\text{cr}}\sim m_\tau\, \E^{-\frac{1}{4\pi b_0 \alpha(m_\tau)}}
\simeq m_\tau\, \E^{-\frac{6\pi}{11\Nc \alpha(m_\tau)}} \left(
1-\epsilon \Nf + \mathcal O ((\epsilon \Nf)^2)\right), 
\end{equation}
where $\epsilon = \frac{12 \pi}{121 \Nc^2 \alpha(m_\tau)} \simeq
0.107$ for $\Nc=3$. This simple estimate hence predicts a linear
decrease of the phase boundary $T_{\text{cr}}(\Nf)$ for small $\Nf$,
as is confirmed by the full solution plotted in Fig.~\ref{tc_nf}.
Actually, this estimate is also quantitatively accurate, since it
predicts a relative difference for $\Tc$ for $\Nf$=2 and 3 flavors of
$\Delta\simeq 0.146$ which is in very good agreement with the full
result, given in Eq. \eqref{eq:Deltares}. We conclude that the shape
of the phase boundary for small $\Nf$ is basically dominated by
fermionic screening.

For larger $\Nf$, the above estimate can no longer be used, because
neither one-loop perturbation theory nor the $\Nf$ expansion are
justified. For values of $\Nf$ close to the critical value
$\Nf^{\text{cr}}$, a different analytic argument can be made: here the
running coupling has to come close to its maximal value in order to be
strong enough to trigger \xsb. The maximal value is, of course,
close to the IR fixed point value $\alpha_\ast$ attained for
$T=0$. Even though at finite $T$ the coupling is eventually governed
by the $3d$ fixed point implying a linear decrease with $k$, the \xsb\
properties will still be dictated by the maximum coupling value, which roughly corresponds 
to the $T=0$ fixed point. In the fixed-point regime, we can
approximate the $\beta_{g^2}$ function by a linear expansion about the
fixed-point value,
\begin{equation}
\beta_{g^2}\equiv \pat g^2 =-\Theta\, (g^2-g_\ast^2)+ \mathcal
O((g^2-g_\ast^2)^2), \label{eq:FPR}
\end{equation}
where the universal ``critical exponent'' $\Theta$ denotes the
(negative) first expansion coefficient. We know that $\Theta<0$, since
the fixed point is IR attractive. For vanishing temperature, we find
an approximate linear dependence of $\Theta$ on $\Nf$, cf.
Tab.~\ref{tab:Theta}.

\begin{table}[]
\begin{center}\begin{tabular}{|c|c|c|c|c|c|c|c|c|c|c|c|}
\hline 
$\Nf$&
0&
4&
5&
6&
7&
8&
9&
10&
11&
12&
13\tabularnewline
\hline
\hline 
$-\Theta$&
6.39&
5.50&
4.99&
4.41&
3.82&
3.19&
2.58&
1.97&
1.42&
0.95&
0.57\tabularnewline
\hline
\end{tabular}\end{center}

\caption{The "critical exponent" $\Theta$ for different values of $\Nf$ for
$T=0$.}
\label{tab:Theta}
\end{table}

The solution of Eq.~\eqref{eq:FPR} for the running coupling in the
fixed-point regime reads
\begin{equation}
g^2(k)=g_\ast^2-\left(\frac{k}{k_0}\right)^{-\Theta}, \label{eq:FPsol}
\end{equation}
where the scale $k_0$ is implicitly defined by a suitable initial
condition (to be set in the fixed-point regime) and is kept fixed 
in the following. It provides for all
dimensionful scales in the following and is related to the initial
$\tau$ mass scale by RG evolution. Our criterion for \xsb\ to occur is
that $g^2(k)$ should exceed $g_{\text{cr}}^2$ for some value of
$k=k_{\text{cr}}$. We expect that this scale $k_{\text{cr}}$ is
generically somewhat larger than the temperature, since for $k$
smaller than $T$ the coupling decreases again owing to the $3d$ fixed
point.\footnote{Indeed, this assumption is justified, since we find in
  the full calculation that $k_{\text{cr}}\gg T$ for large $\Nf$ and
  for temperatures in the vicinity of the critical temperature
  $T_{\text{cr}}$ .}  This allows us to ignore the $T$ dependence of
the running coupling $g^2$ and of the critical coupling
$g_{\text{cr}}$ as a rough approximation, since the $T$ dependence of
the threshold functions is rather weak for $T\lesssim k$. From
Eq.~\eqref{eq:FPsol} and the condition $g^2(k_{\text{cr}})=
g^2_{\text{cr}}$, we derive the estimate
\begin{equation}
k_{\text{cr}}\simeq k_0\, (g_\ast^2
-g_{\text{cr}}^2)^{-\frac{1}{\Theta}}. \label{eq:kcrest}
\end{equation}
This scale $k_{\text{cr}}$ plays the same role as the crossover scale
$k_{\text{co}}$ in the small-$\Nf$ argument given above: it sets the
scale for $T_{\text{cr}}\sim k_{\text{cr}}$, with a proportionality
coefficient provided by the solution of the full flow. To conclude the
argument, we note that the IR fixed-point value $g_\ast^2$ roughly
depends linearly on $\Nf$, since the quark contribution to the
coupling flow $\eta^{\text{q}}$ is linear in $\Nf$. From
Eq.~\eqref{eq:kcrest}, we thus find the relation
\begin{equation}
T_{\text{cr}}\sim k_0 |\Nf -\Nf^{\text{cr}}|^{-\frac{1}{\Theta}},
\label{eq:TcrTheta}
\end{equation}
which is expected to hold near $\Nf^{\text{cr}}$ for
$\Nf\leq\Nf^{\text{cr}}$. Here, $\Theta$ should be evaluated at
$\Nf^{\text{cr}}$.\footnote{Accounting for the $\Nf$ dependence of
$\Theta$ by an expansion around $\Nf^{\text{cr}}$ yields mild
logarithmic corrections to Eq.~\eqref{eq:TcrTheta}.}  Relation
\eqref{eq:TcrTheta} is an analytic prediction for the shape of the
chiral phase boundary in the ($T,\Nf$) plane of QCD. Remarkably, it
relates two universal quantities with each other: the phase boundary
and the IR critical exponent.

This relation can be checked with a fit of the full numerical result
parametrized by the RHS of Eq.~\eqref{eq:TcrTheta}. In fact, the fit
result, $\Theta_{\mathrm{fit}}\simeq -0.60$, determined from the phase
boundary agrees with the direct determination of the critical exponent
from the zero-temperature $\beta$ function,
$\Theta(\Nf^{\text{cr}}\simeq12.9)\simeq -0.60$, within a one-percent
accuracy (cf.~Table \ref{tab:Theta}).  The fit is depicted by the
dashed line in Fig.~\ref{tc_nf}.  In particular, the fact that
$|\Theta|<1$ near $\Nf^{\text{cr}}$ explains the flattening of the
phase boundary near the critical flavor number. 

Qualitatively, relation \eqref{eq:TcrTheta} is a consequence of the
IR fixed-point scenario predicted by our truncated flow equation. We
emphasize, however, that the quantitative results for universal
quantities such as $\Theta$ are likely to be affected by truncation
errors. These can be reduced by an optimization of the present flow;
we expect from preliminary regulator studies that more reliable
estimates of $\Theta$ yield smaller absolute values and, thus, a more
pronounced flattening of the phase boundary. 

We are aware of the fact that the relation \eqref{eq:TcrTheta} is
difficult to test, for instance, by lattice gauge theory: neither the
fixed-point scenario in the deep IR nor large flavor numbers are
easily accessible, even though there are promising investigations that
have collected evidence for the IR fixed-point scenario in the Landau
gauge \cite{Silva:2005hb,Ilgenfritz:2006gp} (see also 
\cite{Cucchieri:1997fy,Bonnet:2001uh,Bogolubsky:2005wf}) as well as the existence of a
critical flavor number \cite{Iwasaki:1991mr}. Given
the conceptual simplicity of the fixed-point scenario in combination
with \xsb, further lattice studies are certainly worthwhile.

\section{Conclusions and outlook}
\label{sec:conc}

We have obtained new nonperturbative results for the chiral phase
boundary of QCD in the plane spanned by temperature and quark flavor
number. Our work is based on the functional RG which provides for a
functional differential formulation of QCD in terms of a flow equation
for the effective action. We have studied this effective action from
first principles in a systematic and consistent operator expansion
which is partly reminiscent to a gradient expansion. We consider the
truncated expansion as a minimal approximation of the effective action
that is capable to access the nonperturbative IR domain and address
the phenomenon of chiral symmetry breaking.

In the gluon sector, this truncation provides for a stable flow of the
gauge coupling, running into a fixed point in the IR at zero
temperature in agreement with the results of \cite{Gies:2002af} for
the pure glue sector. As a new result, we find that the $3d$ analogue
of this IR fixed point governs the flow of the gauge coupling at
finite temperature for scales $k\ll T$.  Our truncation in the quark
sector facilitates a description of critical dynamics with a
gluon-driven approach to \xsb.  The resulting picture for \xsb\ is
comparatively simple: \xsb\ requires the coupling to exceed a critical
value $g_{\text{cr}}$.  Whether or not this critical coupling is
reached depends on the RG flow of the gauge coupling. The IR
fixed-point scenario generically puts an upper bound on the maximal
coupling value which depends on the external parameters such as
temperature and quark flavor number. Of course, the interplay between
the gluon and quark sectors in general, and between gauge coupling and
critical coupling in particular, is highly nonlinear, since both
sectors back-react onto each other in a manner which is quantitatively
captured by the flow equation.

The resulting phase boundary in the ($T,\Nf$) plane exhibits a
characteristic shape which can analytically be understood in terms of
simple physical mechanisms: for small $\Nf$, we observe a linear
decrease of $T_{\text{cr}}$ as a function of $\Nf$; this is a direct
consequence of the charge-screening properties of light
fermions. Also, this screening nature is ultimately responsible for
the existence of a critical flavor number $\Nf^{\text{cr}}$ above
which the system remains in the chirally symmetric phase even at zero
temperature (even though the theory is still asymptotically free for
$\Nf$ not too much larger than $\Nf^{\text{cr}}$).  The shape of the
phase boundary near the critical flavor number, $\Nf\lesssim
\Nf^{\text{cr}}$, is most interesting from our viewpoint. In this
region, the critical temperature is very small, and thus the system is
probed in the deep IR.  As a main result of this paper, we have shown
that this connection becomes most obvious in an intriguing relation
between the shape of the phase boundary for $\Nf\lesssim
\Nf^{\text{cr}}$ and the IR critical exponent $\Theta$ of the running
coupling at zero temperature.  In particular, the flattening of the
phase boundary in this regime is a direct consequence of $|\Theta|$
being smaller than 1. Since both the shape of the phase boundary and
the critical exponent are universal quantities, their relation is a
generic prediction of our analysis. It can directly be tested by other
nonperturbative methods, even though it may numerically be expensive,
e.g., in lattice simulations.

Let us now critically assess the reliability of our results.
Truncating the effective action, at first sight, is an uncontrolled
approximation which can a priori be justified only with some insight
into the physical mechanisms. The truncation in the quark sector
supporting potential critical dynamics is an obvious example for this.
The approximation can become (more) controlled if the inclusion of
higher-order operators does not lead to serious modifications of the
results. In the quark sector, it can indeed easily be verified that
the contribution of many higher-order operators such as $(\yb\psi)^4$
or mixed gluonic-fermionic operators is generically suppressed by the
one-loop structure of the flow equation or the fixed-point argument
given below Eq.~\eqref{betaeq}. This holds at least in the symmetric
regime, which is sufficient to trace out the phase boundary. By
contrast, we are not aware of similar arguments for the gluonic
sector; here, higher-order expansions involving, e.g., $(F_{\mu\nu}
\widetilde{F}^{\mu\nu})^2$ or operators with covariant derivatives or
ghost fields eventually have to be used to verify the expansion
scheme. At finite temperature, the difference between so-called
electric and magnetic sectors can become important, as mediated by
operators involving the heat-bath four-velocity $u_{\mu}$, e.g.,
$(F_{\mu\nu} u_\nu)^2$.  In view of results obtained in the Landau
gauge \cite{vonSmekal:1997is}, the inclusion of ghost contributions in
the gauge sector appears important if not mandatory for a description
of color confinement.  A posteriori, the truncation can be verified by
a direct comparison with lattice results. In the present case, this
cross-check shows satisfactory agreement.

The stability of the present
results can also be studied by varying the regulator. Since universal
quantities are independent of the regulator in the exact theory, any
such regulator dependence of the truncated system is a measure for the
reliability of the truncation.  As was already quantitatively
verified at vanishing temperature in \cite{Gies:2005as}, the present
quark sector shows surprisingly little dependence on the regulator
which strongly supports the truncation.  By contrast, we do not expect
such a regulator independence to hold in the truncated gluonic sector.
If so, it is advisable to improve results for universal quantities
towards their physical values. This can indeed be done by using
stability criteria for the flow equation which have lead to
optimization schemes
\cite{Litim:2001up,Pawlowski:2005xe,Litim:2005us}.  We
  expect that the use of such optimized regulators give better results
  for dimensionless quantities, e.g. Eq.  \eqref{eq:Deltares} or the
  IR critical exponent $\Theta$.  In any case, we have confirmed
that, for instance, the linear regulator \cite{Litim:2001up},
which satisfies optimization criteria in various systems, leads to the
same qualitative results as presented above. Further regulator studies
are left to future work.

Further generalizations of our work will aim at a quantitative study
of the effect of finite quark masses; the formalism of which has
largely been developed already in this work. Owing to the mechanism of
fermionic decoupling, we expect that the largest modifications arise
from a realistic strange quark mass which is of the order of the
characteristic scales such as $T_{\text{cr}}$ or the scale of \xsb. 

Let us finally stress that our whole quantitative analysis relies on
only one physical input parameter, namely the value of the gauge
coupling at a physical input scale. This clearly demonstrates the
predictive power of the functional RG approach for full QCD, and
serves as a promising starting point for further phenomenological
applications.

\section*{Acknowledgment}

The authors are grateful to J.~Jaeckel, J.M. Pawlowski, and
H.-J.~Pirner for useful discussions.  H.G.~ acknowledges support by
the DFG under contract Gi 328/1-3 (Emmy-Noether
program). J.B. acknowledges support by the GSI Darmstadt.

\begin{appendix}

\renewcommand{\thesection}{\mbox{\Alph{section}}}
\renewcommand{\theequation}{\mbox{\Alph{section}.\arabic{equation}}}
\setcounter{section}{0}
\setcounter{equation}{0}

\section{Thermal moments and threshold functions}
\label{sec:Regulator_function}

\setcounter{equation}{0}

\subsection{Thermal moments}
\label{sec:moments}

Let us first define the auxiliary functions $f$ which are first
introduced in Eq.~\eqref{eq:fe_wpot}:
\begin{eqnarray}
f_{T}^{A}(u,v) & = &
2\sqrt{4\pi}v\sum_{q=-\infty}^{\infty}\int_{0}^{\infty}dx\,\E^{-(2\pi
  vx)^{2}u}\cos(2\pi qx)\,,\label{eq:fT_F}\\ 
f_{T}^{\psi}(u,v) & = & 2\sqrt{4\pi}v\sum_{q=-\infty}^{\infty}(-1)^{q}
  \int_{0}^{\infty}dx\,\E^{-(2\pi
  vx)^{2}u}\cos(2\pi qx)\,,\label{eq:fT_psi}\\ 
f^{\psi}(u) & = & \frac{1}{2}\frac{1}{u^{e_{d}}}u\coth(u),\\
f_{1}^{A}(u) & = & \frac{1}{u^{e_{d}}}\Big(e_{d}\,\frac{u}{\sinh
  u}+2u\sinh u\Big)\,,\\ 
f_{2}^{A}(u) & = & \frac{1}{2}\frac{1}{u^{e_{d}}}\frac{u}{\sinh u}\,,\\
f_{3}^{A}(u) & = & \frac{1}{u^{e_{d}}}(1-v)\,,\\
f_{4}^{A}(u,v) & = &
2\sqrt{4\pi}v\sum_{q=-\infty}^{\infty}\int_{0}^{\infty}dx(2\pi
vx)^{d-1}
\Gamma\big(\!-\! e_{d},(2\pi vx)^{2}u\big)\cos(2\pi qx).\label{eq:f4_F}
\end{eqnarray}
{Here, the sum over $q$ arises from the application of 
Poisson's Formula to the (usual) Matsubara sum.}
These functions are needed for the construction of the thermal moments
$\bar{h}_{j}^{\psi}$ $\bar{h}_{j}^{A}$, $\bar{g}_{j}^{A}$,
$\bar{H}_{j}^{A}$ and $\bar{G}_{j}^{A}$ which are related to the
regulator via Eqs.~\eqref{eq:def_h}, \eqref{eq:def_g} and
\eqref{eq:def_hpsi},\eqref{eq:hprop} by
\begin{equation}
\bar{h}_{j}^{\psi}:=\bar{h}_{j}^{\psi}\big(\tilde{m},v\big)=\int_{0}^{\infty}\! ds\,\widetilde{h}^{\psi}(s,\tilde{m})s^{j}f_{T}^{\psi}\big(s,v\big)\,,\label{eq:hbar_psi1}
\end{equation}
\begin{equation}
\bar{h}_{j}^{A}:=\bar{h}_{j}^{A}\big(v\big)=\int_{0}^{\infty}\! ds\,\widetilde{h}(s)s^{j}f_{T}^{A}\big(s,v\big)\,,\label{eq:hbar1}
\end{equation}
\begin{equation}
\bar{g}_{j}^{A}:=\bar{g}_{j}^{A}\big(v\big)=\int_{0}^{\infty}\! ds\,\widetilde{g}(s)s^{j}f_{T}^{A}\big(s,v\big)\,,\label{eq:gbar1}
\end{equation}
\begin{equation}
\bar{H}_{j}^{A}:=\bar{H}_{j}^{A}\big(v\big)=\int_{0}^{\infty}\! ds\,\widetilde{h}(s)s^{j}f_{4}^{A}\big(s,v\big)\,,\label{eq:Hbar1}
\end{equation}
\begin{equation}
\bar{G}_{j}^{A}:=\bar{G}_{j}^{A}\big(v\big)=\int_{0}^{\infty}\! ds\,\widetilde{g}(s)s^{j}f_{4}^{A}\big(s,v\big)\,,\label{eq:Gbar1}
\end{equation}
where $\tilde{m}$ denotes a dimensionless quark mass parameter. It is
more convenient to express the moments in terms of the regulator
functions $h(y)$ and $g(y)$ in momentum space which are defined in
Eqs. \eqref{eq:def_h} and \eqref{eq:def_g}. In order to obtain the
representations for $\bar{h}_{j}$ and $\bar{g}_{j}$, we introduce
\begin{equation}
\frac{s^{b+1}}{\Gamma(b+1)}\int_{0}^{\infty}du\, u^{b}\E^{-su}=1\qquad(b>-1)\label{eq:gamma_u}
\end{equation}
in Eq. \eqref{eq:hbar1}, \eqref{eq:gbar1} and \eqref{eq:hbar_psi1}
and use Eq. \eqref{eq:fT_F} and \eqref{eq:fT_psi}, respectively:
\begin{equation}
\bar{h}_{j}^{\psi}=\frac{2}{\Gamma(b+1)\sqrt{\pi}}\sum_{q=-\infty}^{\infty}(-1)^{q}\int_{0}^{\infty}\! dx\,\cos\Big(q\frac{x}{v}\Big)\left.\Big(\!-\!\frac{d}{dy}\Big)^{j+b+1}\int_{0}^{\infty}\! du\, u^{b}h^{\psi}(y+u+x^{2},\tilde{m})\right|_{y=0} ,\label{eq:hbar_psi2}
\end{equation}
\begin{equation}
\bar{h}_{j}^{A}=\frac{2}{\Gamma(b+1)\sqrt{\pi}}\sum_{q=-\infty}^{\infty}\int_{0}^{\infty}\! dx\,\cos\Big(q\frac{x}{v}\Big)\left.\Big(\!-\!\frac{d}{dy}\Big)^{j+b+1}\int_{0}^{\infty}\! du\, 
u^{b}h(y+u+x^{2})\right|_{y=0} ,\label{eq:hbar2}
\end{equation}
\begin{equation}
\bar{g}_{j}^{A}=\frac{2}{\Gamma(b+1)\sqrt{\pi}}\sum_{q=-\infty}^{\infty}\int_{0}^{\infty}\! dx\,\cos\Big(q\frac{x}{v}\Big)\left.\Big(\!-\!\frac{d}{dy}\Big)^{j+b+1}
\int_{0}^{\infty}\! du\, u^{b}g(y+u+x^{2})\right|_{y=0} .\label{eq:gbar2}
\end{equation}
Note that $b$ is an arbritrary parameter which can, e.g., be used
to avoid fractional derivatives. 
By applying Poisson's formula to the (usual) Matsubara sum, we 
have obtained the sum over $q$ which converges fast for $k\gtrsim T$.
Moreover, we need
$\bar{H}_{j}^{A}$ and $\bar{G}_{j}^{A}$ for $j=0$, which are used
in App.  \ref{sec:Resummation}. Integrating Eq. \eqref{eq:Hbar1} and
\eqref{eq:Gbar1} by parts and using Eq. \eqref{eq:gamma_u} and
\eqref{eq:f4_F}, we obtain
\begin{equation}
\bar{H}_{0}^{A}=\frac{2}{\Gamma(b+1)\sqrt{\pi}}\sum_{q=-\infty}^{\infty}\int_{0}^{\infty}\! dx\,\cos\Big(q\frac{x}{v}\Big)\,\left.\Big(-\frac{d}{dy}\Big)^{b-e_{d}}\int_{0}^{\infty}\! du\,\frac{u^{b}}{u+x^{2}}h(y+u+x^{2})\right|_{y=0}\,,\label{eq:Hbar2}
\end{equation}
\begin{equation}
\bar{G}_{0}^{A}=\frac{2}{\Gamma(b+1)\sqrt{\pi}}\sum_{q=-\infty}^{\infty}\int_{0}^{\infty}\! dx\,\cos\Big(q\frac{x}{v}\Big)\,\left.\Big(-\frac{d}{dy}\Big)^{b-e_{d}}\int_{0}^{\infty}\! du\,\frac{u^{b}}{u+x^{2}}g(y+u+x^{2})\right|_{y=0}\,.\label{eq:Gbar2}
\end{equation}
In this paper, we use the exponential regulator. For the gluon and
ghost fields, this regulator is given by
\begin{equation}
R_{k}(\Delta)=\Delta\, r\big({\textstyle \frac{\Delta}{k^{2}}}\big)\quad\mathrm{{with}\quad\mathnormal{r(y)=\frac{1}{\E^{y}-1}}}\,,
\label{reg_A}
\end{equation}
and the functions $h(y)$ and $g(y)$ read \cite{Gies:2002af}
\begin{equation}
\label{eq:bos_hg}
h(y)=\frac{y}{\E^{y}-1}\quad\mathrm{and}\quad g(y)=\E^{-y}\,.
\end{equation}
For the quark fields, the exponential regulator reads
\begin{equation}
\label{reg_psi}
R_{k}^{\psi}(\I\slash{\bar D})=\I\slash{\bar D}\,
r_{\psi}\Big({\textstyle \frac{(\I\slash{\bar
D})^{2}}{k^{2}}}\Big)\quad\mathrm{with}\quad
r_{\psi}(y)=\frac{1}{\sqrt{1-\E^{-y}}}-1\,,
\end{equation}
and the functions $h^{\psi}(y,\frac{m}{k})$ and $g^{\psi}(y,\frac{m}{k})$
are given by
\begin{equation}
\label{eq:quark_hg}
h^{\psi}(y,\tilde{m})=\frac{y^{2}}{(\E^{y}-1)(y+\tilde{m}^{2}(1-\E^{-y}))}\quad\mathrm{and}\quad g^{\psi}(y,\tilde{m})=\frac{y(1-\E^{-y})(1-\sqrt{1-\E^{-y}})}{y+\tilde{m}^{2}(1-\E^{-y})}\,.
\end{equation}
Inserting Eq.~\eqref{eq:bos_hg} and \eqref{eq:quark_hg} into 
Eqs.~\eqref{eq:hbar_psi2}-\eqref{eq:Gbar2} completely determines the
desired thermal moments. 

\subsection{Threshold functions}
\label{sec:thres_fcts}
In Sec. \ref{sec:quarks}, the regulator dependence of the flow
equations of the four-fermion interactions is controlled by threshold
functions. The purely fermionic threshold functions are defined by
\begin{equation}
\label{quark_thres}
l_{n}^{(F)d}(t,w)=n\frac{v_{d-1}}{v_{d}} \,t \sum _{n=-\infty} ^{\infty}
\int _{0} ^{\infty}\, d y y^{\frac{d-3}{2}}\,
\frac{p_{\psi}(y_{\psi})-y_{\psi}
\dot{p}_{\psi}(y_{\psi})}{[p_{\psi}(y_{\psi}) +w]^{n+1}}\, ,
\end{equation}
where $t\equiv T/k$ and $w$ are dimensionless quantities, 
the latter being associated with finite quark masses. Dots
denote derivatives with respect to $y_{\psi}$.  The dimensionless
momentum $y_{\psi}=\tilde{\nu}_{n}^2+y$ depends on the (dimensionless)
fermionic Matsubara frequencies $\tilde{\nu}_{n}=(2n+1)\pi t$.  The
function $p_{\psi}(y_{\psi})$ is related to the regulator shape
function $r_{\psi}$ by
\begin{equation}
r_{\psi}(y_{\psi})=\sqrt{\frac{p_{\psi}(y_{\psi})}{y_{\psi}}}-1\,.
\end{equation}
The factor $v_d^{-1}$ is proportional to the volume of 
the $d$ dimensional unit ball: 
\begin{equation}
v_{d}^{-1}=2^{d+1}\pi ^{\frac{d}{2}}\Gamma\left(\frac{d}{2}\right)\,.
\end{equation}
In Sec. {\ref{sec:quarks}}, we only need $l_{1}^{(F)}$.  Using the
exponential regulator Eq.~\eqref{reg_psi} and $w=0$ for massless
quarks, the fermionic threshold function $l_{1}^{(F)}(t,0)$ reads,
\begin{equation}
l_{1}^{(F)}(t,0)=\sum _{n=-\infty} ^{\infty} (-1)^{n}\E ^{-\frac{n}{2t}}
\, ,\qquad l_{1}^{(F)}(t\rightarrow 0,0) \longrightarrow 1\,.
\label{eq:lF1}
\end{equation}
The threshold functions $l_{n_1,n_2}^{(FB)d}(t,w_1,w_2)$ 
arise from Feynman graphs, incorporating fermionic and 
bosonic fields:
\begin{eqnarray}
l_{n_1,n_2}^{(FB)d}(t,w_1,w_2) & = & \frac{v_{d-1}}{v_{d}} \,t 
\sum _{n=-\infty} ^{\infty}\int _{0} ^{\infty}\, d y y^{\frac{d-3}{2}}\,
\frac{1}{[p_{\psi}(y_{\psi})+w_1]^{n_1}[p_{A}(y_{A})+w_2]^{n_2}}\nn\\
& & \times \left\{\frac{n_1 [p_{\psi}(y_{\psi})-y_{\psi}\dot{p}_{\psi}(y_{\psi})]}{p_{\psi}(y_{\psi}) + w_1} +
\frac{n_2 [p_{A}(y_{A})-y_A\dot{p}_{A}(y_{A})]}{p_{A}(y_{A}) + w_2}\right\}\,.
\end{eqnarray}
Here, $w_1$ and $w_2$ are dimensionless arguments and dots denote
derivatives with respect to $y_{\psi}$ and $y_{A}$, respectively.  In
analogy to the fermionic case, the dimensionless bosonic momentum
$y_{A}=\tilde{\omega}_{n}^2+y$ depends on the (dimensionless) bosonic
Matsubara frequencies $\tilde{\omega}_{n}^2=4\pi^2 n^2 t^2$.  The
(bosonic) regulator shape function $r$ is connected with $p_{A}$ by
the relation
\begin{equation}
p_{A}(y_{A})=y_{A}[1+r(y_{A})]\,.
\end{equation}
In Sect. \ref{sec:quarks}, we need $l_{1,1}^{(FB)4}$ and
$l_{1,2}^{(FB)4}$. Using the exponential regulator
Eq.~\eqref{reg_A}~and~\eqref{reg_psi}, we can calculate the
integrals analytically in the limit $t\rightarrow 0$ and
$w_1=w_2=0$ for $d=4$:
\begin{equation}
\lim_{t\rightarrow0}l_{1,1}^{(FB)4}(t,0,0)=1\qquad
\mathrm{and}\qquad
\lim_{t\rightarrow0}l_{1,2}^{(FB)4}(t,0,0)=3\ln({\textstyle\frac{4}{3}})\,.
\label{eq:lFB}
\end{equation}
For $t\rightarrow\infty$ or $w\rightarrow\infty$, the
threshold functions $l_{n}^{(F)d}$ and $l_{n_1,n_2}^{(FB)d}$ approach
zero. For finite $t$ and $w$, the threshold functions can easily be evaluated 
numerically.

\section{Resummation of the anomalous dimension}
\label{sec:Resummation}

\setcounter{equation}{0}
Here, we present details for the resummation of the series expansion of 
the anomalous dimension $\eta$,
\begin{equation}
\eta\simeq\sum_{i=0}^{\infty}a^{\text{l.g.}}_{m}G^{m}\,.
\label{eq:sec2_2_eta_series2a} 
\end{equation}
The leading growth (l.g.) coefficients $a_{m}^{l.g.}$ read
\begin{align}
a_{m}^{l.g.} & =a_{m}^{A}+a_{m}^{\text{q}}=4(-2c_{1})^{m-1}\frac{\Gamma(z_{d}+m)\Gamma(m+1)}{\Gamma(z_{d}+1)}\Big[\bar{h}_{2m-e_{d}}^{A}({\textstyle\frac{T}{k}})
(d\!-\!2)\frac{2^{2m}-2}{(2m)!}\tau_{m}^{A}B_{2m}\nonumber \\
 &
 \qquad\qquad\qquad\qquad -\frac{4}{\Gamma(2m)}\tau_{m}^{A}
 \bar{h}_{2m-e_{d}}^{A}({\textstyle\frac{T}{k}})
+4^{m+1}\frac{B_{2m}}{(2m)!}\tau_{m}^{\psi}
\sum_{i=1}^{\Nf}\bar{h}_{2m-e_{d}}^{\psi}({\textstyle
 \frac{m_{i}}{k}},{\textstyle
 \frac{T}{k}})\Big]\,,\label{eq:app_res_1} 
\end{align}

where $B_{2m}$ are the Bernoulli numbers and $z_{d}$ is defined
as
\begin{equation}
z_{d}:=(d-1)(\Nc^{2}-1)c_{2}\,.\label{eq:app_res_def_zd}
\end{equation}
The temperature and regulator-dependent functions $c_{1}$ and $c_{2}$
are given by
\begin{eqnarray}
c_{1} & = & 2\big(\bar{H}_{0}^{A}\big({\textstyle \frac{T}{k}}\big)-\bar{G}_{0}^{A}\big({\textstyle \frac{T}{k}}\big)\big)\,,\label{eq:app_res_def_c1}\\
c_{2} & = & \frac{\bar{h}_{-e_{d}}^{A}\big(\frac{T}{k}\big)-
\bar{g}_{-e_{d}}^{A}\big(\frac{T}{k}\big)}{c_{1}}\,.\label{eq:app_res_def_c2}
\end{eqnarray}
Note that $c_{1}>0$ and $c_{2}>0$ for $\frac{T}{k}\geq0$. In the
limits $\frac{T}{k}\rightarrow0$ and $\frac{T}{k}\rightarrow\infty$,
$c_{1}$ and $c_{2}$ are given by
\begin{eqnarray}
\lim_{\frac{T}{k}\rightarrow0}c_{1} & = & c_{1} ^{0}=\frac{4}{d}\Big(\frac{d}{2}\zeta\Big(1+\frac{d}{2}\Big)-1\Big)\,,\label{eq:app_res_c1_t0}\\
\lim_{\frac{T}{k}\rightarrow0}c_{2} & = & c_{2}^{0}=\frac{d}{4}\,,\label{eq:app_res_c2_t0}\\
\lim_{\frac{T}{k}\rightarrow\infty}{\textstyle{\frac{k}{T}}}c_{1} & = & {c}_{1}^{\infty}=2\sqrt{4\pi}\Big(\zeta\Big(1+e_{d}\Big)-\frac{2}{d-1}\Big)\,,\label{eq:app_res_c1_tinf}\\
\lim_{\frac{T}{k}\rightarrow\infty}c_{2} & = & c_{2}^{\infty}=\frac{e_{d}\zeta(1+e_{d})-1}{2(\zeta(1+e_{d})-\frac{2}{d-1})}\,,\label{eq:app_res_c2_tinf}
\end{eqnarray}
where we have used Eq. \eqref{eq:hbar1}-\eqref{eq:Gbar2} for the exponential 
regulator and $\zeta(x)$ denotes the Riemann Zeta function.

Now, we perform the resummation of $\eta$ along the lines of \cite{Gies:2002af}:
We split the anomalous dimension Eq. \eqref{eq:sec2_2_eta_series}
into three contributions,
\begin{equation}
\eta=\eta_{1}^{A}+\eta_{2}^{A}+\eta^{\text{q}}\,,
\end{equation}
where $\eta_{1}^{A}$ corresponds to the resummation of the term $\sim\tau_{m}^{A}B_{2m}$
in Eq. (\ref{eq:app_res_1}) and $\eta_{2}^{A}$ to the resummation
of the term containing the Nielsen-Olesen unstable mode ($\sim1/\Gamma(2m)$)
, representing the leading and subleading growth, respectively. The
remaining contributions are contained in $\eta^{\text{q}}$. 

First, we confine ourselves to $SU(N_c =2)$ for which the group
theoretical factors are $\tau_{m}^{A}~=~N_c$ and
$\tau_{m}^{\psi}~=~\Nc\,(1/4)^{m}~=~2\,(1/4)^{m}$ (see Appendix
\ref{sec:color} for details), but we artificially retain the $\Nc$
dependence in all terms in order to simplify the generalization to
gauge groups of higher rank.

We start with the resummation of $\eta_{1}^{A}$: For this purpose,
we use the standard integral representation of the $\Gamma$ functions
\cite{Grad:2000gr},
\begin{equation}
\Gamma(z_{d}+m)\Gamma(m+1)=\int_{0}^{\infty}\! ds_{1}\int_{0}^{\infty}\! ds_{2}\, s_{1}s_{2}^{z_{d}}(s_{1}s_{2})^{m-1}\E^{-(s_{1}+s_{2})}=\int_{0}^{\infty}\! dp\,\tilde{K}_{z_{d}-1}(p)\, p^{m-1},\label{eq:gamma_bess}
\end{equation}
where we have introduced the modified Bessel function
\begin{equation}
\tilde{K}_{z_{d}-1}(s)=2s^{\frac{1}{2}(z_{d}+1)}K_{z_{d}-1}(2\sqrt{s})\,.
\end{equation}
Furthermore, we use the series representation of the Bernoulli numbers
\cite{Grad:2000gr},
\begin{equation}
\frac{B_{2m}}{(2m)!}=2\frac{(-1)^{m-1}}{(2\pi)^{2m}}\sum_{l=1}^{\infty}\,\frac{1}{l^{2m}}\,.\label{eq:app_res_Bern}
\end{equation}
With the aid of Eq. \eqref{eq:gamma_bess} and \eqref{eq:app_res_Bern},
we rewrite $\eta_{1}^{A}$ as follows
\begin{equation}
\eta_{1}^{A}=\frac{4(d\!-\!2)\Nc G}{\pi^{2}\Gamma(z_{d}\!+\!1)}\,\sum_{m=1}^{\infty}\sum_{l=1}^{\infty}\frac{1}{l^{2}}\int_{0}^{\infty}\! dp\,\tilde{K}_{z_{d}-1}(p)\,\bar{h}_{2m-e_{d}}^{A}({\textstyle\frac{T}{k}})\Big[2\Big(\frac{2Gpc_{1}}{\pi^{2}l^{2}}\Big)^{m-1}\!-\Big(\frac{Gpc_{1}}{2\pi^{2}l^{2}}\Big)^{m-1}\Big]\,.\label{eq:app_etaa1}
\end{equation}
In order to perform the summation over $m$, we define
\begin{align}
S_{b}^{A}(q,v)= & \sum_{l=1}^{\infty}\frac{1}{l^{2}}\sum_{m=1}^{\infty}\Big(\frac{q}{l^{2}}\Big)^{m-1}\bar{h}_{2m-e_{d}}^{A}(v)\nonumber \\
= & \frac{2}{\sqrt{\pi}}\sum_{l=1}^{\infty}\sum_{m=0}^{\infty}\sum_{n=-\infty}^{\infty}\int_{0}^{\infty}\! dx\cos\Big(\frac{nx}{v}\Big)\int_{0}^{\infty}\! dt\,\frac{\E^{-t}}{l^{2}}\int_{0}^{\infty}\! ds\,\tilde{h}(s)\frac{s^{2-e_{d}}}{(2m)!}\Big(\frac{st\sqrt{q}}{l}\Big)^{2m}\E^{-sx^{2}}\nonumber \\
= & \frac{1}{\Gamma(b\!+\!1)\sqrt{q\pi}}\sum_{n=-\infty}^{\infty}\int_{0}^{\infty}\! dx\,\cos\Big(\frac{nx}{v}\Big)\int_{0}^{\infty}\! dt\,\Li_{1}\Big(\E^{-\frac{t}{\sqrt{q}}}\Big)\,\sigma_{b}^{A}(x^{2},t)\,,\label{eq:def_SF}
\end{align}
where we have used Eqs. \eqref{eq:hbar1}, \eqref{eq:fT_F} and \eqref{eq:gamma_u}.
The auxiliary function $\sigma_{b}^{A}$ is defined as
\begin{equation}
\sigma_{b}^{A}(x,t)=\Big(\!-\!\frac{d}{dy}\Big)^{b+3-e_{d}}\left.\int_{0}^{\infty}\! du\, u^{b}\Big[h(y\!+\! u\!+\! x^{2}\!-\! t)+h(y\!+\! u\!+\! x^{2}\!+\! t)\Big]\right|_{y=0}\,.\label{eq:def_sigma}
\end{equation}
Using Eq. \eqref{eq:def_SF}, we obtain the final expression for $\eta_{1}^{A}$,
\begin{equation}
\eta_{1}^{A}=\frac{4(d\!-\!2)\Nc G}{\pi^{2}\Gamma(z_{d}\!+\!1)}\,
 \int_{0}^{\infty}\! dp\,\tilde{K}_{z_{d}-1}(p)\,
 \Big[2S_{b}^{A}\Big(\frac{2Gpc_{1}}{\pi^{2}},\frac{T}{k}\Big)\!-\!
  S_{b}^{A}\Big(\frac{Gpc_{1}}{2\pi^{2}},\frac{T}{k}\Big)\Big]
  \, ,\label{eq:app_etaa2}
\end{equation}
which can straightforwardly be evaluated numerically.

Now we turn to the calculation of $\eta_{2}^{A}$, the
subleading-growth part of $\eta$.  Here, a careful treatment of the
zeroth Matsubara frequency which contains the Nielsen-Olesen mode, is
necessary. More specifically, we transform the modified moments
$\bar{h}_{j}^{A}$ in Eq. \eqref{eq:hbar2} into a sum over Matsubara
frequencies and insert a regulator function
$\mathcal{P}\big({\textstyle \frac{T}{k}}\big)$ for the unstable mode,
\begin{equation}
\bar{h}_{j}^{A,reg}(v)=\sqrt{4\pi}v\sum_{n=-\infty}^{\infty}\int_{0}^{\infty}\! ds\,\widetilde{h}(s)s^{j}\E^{-s\widetilde{\mathcal{P}}_{n}(v)}\,.\label{eq:h_reg}
\end{equation}
\\
Here, we have introduced
\begin{equation}
\widetilde{\mathcal{P}}_{n}(v)=\left\{ \begin{array}{c}
(2\pi nv)^{2}\quad(n\neq0)\\
\mathcal{P}(v)\,\quad\quad(n=0)\end{array}\right.\,.
\end{equation}
The function $\mathcal{P}(v)$ specifies the regularization 
of the Nielsen-Olesen mode and is defined in Eq. \eqref{eq:no_reg}; 
the other modes with $n \neq 0$ remain unmodified.

We rewrite $\eta_{2}^{A}$ by means of Eq. \eqref{eq:gamma_bess},
\begin{equation}
\eta_{2}^{A}=-\frac{16\Nc G}{\Gamma(z_{d}\!+\!1)}\,\sum_{m=1}^{\infty}\frac{1}{\Gamma(2m)}\int_{0}^{\infty}\! dp\,\tilde{K}_{z_{d}-1}(p)\,\bar{h}_{2m-e_{d}}^{A,reg}({\textstyle\frac{T}{k}})\Big(-2Gpc_{1}\Big)^{m-1}\,.\label{eq:app_etab1}
\end{equation}
\\
Now it is convenient to introduce an auxiliary function $T^{A}(q)$
which is defined as
\begin{align}
T_{b}^{A}(q,v)= & \sum_{m=1}^{\infty}\frac{1}{\Gamma(2m)}\Big(-q\Big)^{m-1}\bar{h}_{2m-e_{d}}^{A,reg}(v)\nonumber \\
= & \frac{\sqrt{\pi}v}{\Gamma(b+1)}\sum_{n=-\infty}^{\infty} \int_{0}^{1}\! dt \int_{0}^{\infty}\! du\,
 u^{b}\int_{0}^{\infty}\!ds\,\tilde{h}(s)s^{b+3-e_{d}}
 \E^{-s(u+\widetilde{\mathcal{P}}_{n}(v))}\Big[ \E^{-s t\sqrt{-q}}
 + \E^{s t\sqrt{-q}}\Big]\nonumber \\
= & \frac{\sqrt{\pi}v}{\Gamma(b\!+\!1)}\sum_{n=-\infty}^{\infty}\, 
\vartheta_{b}^{A}(\widetilde{\mathcal{P}}_{n}(v),q)\,,\label{eq:def_TF}
\end{align}
Here, we have used Eqs. \eqref{eq:h_reg} and \eqref{eq:gamma_u}.
Furthermore, we have defined the function $\vartheta_{b}^{A}$:
\begin{equation}
\vartheta_{b}^{A}(x,q)=\left.\Big(\!-\!\frac{d}{dy}\Big)^{b+3-e_{d}}
\int_{0}^{1}\! dt\int_{0}^{\infty}\! du\, u^{b}
\Big[h(y\!+\! u\!+x\!-\! t\sqrt{-q})+h(y\!+\! u\!+\!x+\! t\sqrt{-q})\Big]\right|_{y=0}\,.\label{eq:def_theta}
\end{equation}
Applying Eq. \eqref{eq:def_TF} to Eq. \eqref{eq:app_etab1}, we obtain
\begin{equation}
\eta_{2}^{A}=-\frac{16\Nc G}{\Gamma(z_{d}\!+\!1)}\,\int_{0}^{\infty}\! 
dp\,\tilde{K}_{z_{d}-1}(p)\, T_{b}^{A}(2Gpc_{1},{\textstyle \frac{T}{k}})\, ,\label{eq:app_etab2}
\end{equation}
which can straightforwardly be evaluated numerically.
Finally, we have to calculate the contribution of the quarks to the gluon 
anomalous dimension. Performing analogous steps along the lines of the
calculation of $\eta_{1}^{A}$, we obtain
\begin{equation}
\eta^{\text{q}}=\frac{8\Nc G}{\pi^{2}\Gamma(z_{d}\!+\!1)}
\sum_{i=1}^{\Nf}\int_{0}^{\infty}\! dp\,\tilde{K}_{z_{d}-1}(p)\,
 S_{b}^{\psi}\Big(\frac{pGc_{1}}{2\pi^{2}},\frac{T}{k},\frac{m_{i}}{k}\Big)\,.
\label{eq:eta_quark2}
\end{equation}
The auxiliary function $S_{b}^{\psi}(q,\tilde{m})$ is defined as
\begin{equation}
S_{b}^{\psi}(q,v,\tilde{m})=\frac{1}{\Gamma(b\!+\!1)\sqrt{4\pi q}}\sum_{n=-\infty}^{\infty}(-1)^{n}\int_{0}^{\infty}\! dx\,\cos\Big(\frac{nx}{v}\Big)\int_{0}^{\infty}\! dt\,\Li_{1}\Big(\E^{-\frac{t}{\sqrt{q}}}\Big)\,\sigma_{b}^{\psi}(u,x^{2},t,\tilde{m})\,,\label{eq:def_SF_psi}
\end{equation}
where $\sigma_{b}^{\psi}(u,x,t,\tilde{m})$ is given by
\begin{eqnarray}
\sigma_{b}^{\psi}(u,x,t,\tilde{m}) & = & \Big(\!-\!\frac{d}{dy}\Big)^{b+3-e_{d}}\int_{0}^{\infty}\! du\, u^{b}\Big[h_{s}^{\psi}\Big(\sqrt{y\!+\! u\!+\! x\!-\! t},\tilde{m}\Big)
+h_{s}^{\psi}\Big(\sqrt{y\!+\! u\!+\! x\!+\! t},\tilde{m}\Big)\nonumber \\
 &  & \qquad\quad\left.+h_{s}^{\psi}\Big(\!-\!\sqrt{y\!+\! u\!+\! x\!-\! t},\tilde{m}\Big)
+h_{s}^{\psi}\Big(\!-\!\sqrt{y\!+\! u\!+\! x\!+\! t},\tilde{m}\Big)\Big]\right|_{y=0}\,.\label{eq:def_sigma_psi}
\end{eqnarray}
The regulator function occurs in the function $h_{s}^{\psi}(\sqrt{y},\tilde{m})$ which is related 
to $h^{\psi}(y,\tilde{m})$ by
\begin{equation}
h_{s}^{\psi}(\sqrt{y},\tilde{m})\equiv h^{\psi}(y,\tilde{m})\,.
\end{equation}
There is one essential difference between the resummation of
$\eta_{1/2}^{A}$ and that of $\eta ^{q}$: the regulator shape function
$r(y)$ can be expanded in powers of $y$, while the corresponding
function $r_{\psi}(y)$ for the quark fields should have a power series
in $\sqrt{y}$ which is a consequence of chiral symmetry
\cite{Jungnickel:1995fp}; this explains the notation
$h_{s}^{\psi}(\sqrt{y},\tilde{m})$.

We stress that all integral representations in Eqs.
\eqref{eq:app_etaa2}, \eqref{eq:app_etab2} and \eqref{eq:eta_quark2}
are finite and can be evaluated numerically. For $d=4$ and in the
limit $T\rightarrow0$, the results agree with those of Ref.
\cite{Gies:2002af}.

The remainder of this section deals with a generalization to
higher gauge groups. Since we do not have the explicit
representation of the color factors $\tau_{m}^{A/\psi}$ for gauge groups with
$\Nc\geq3$ at hand, we have to scan the Cartan subalgebra for the
extremal values of $\tau_{m}^{A}$ and $\tau_{m}^{\psi}$. However, as
discussed in App.~\ref{sec:color}, these extremal values of
$\tau_{m}^{A}$ and $\tau_{m}^{\psi}$ can be calculated
straightforwardly. Their insertion into Eq. \eqref{eq:app_res_1} allows
to display the anomalous dimension for $SU(3)$ in terms of the already
calculated formulas for $SU(2)$:
\begin{eqnarray}
\eta_{3}^{\mathrm{{SU(3)}}}
 & = &
 \frac{2}{3}\Big[\eta_{1}^{A}+\eta_{2}^{A}\Big]_{\Nc\rightarrow3} 
 + \frac{1}{3}\Big[\eta_{1}^{A}+\eta_{2}^{A}
   \Big]_{\Nc\rightarrow3,c_{1}\rightarrow c_{1}/4} 
 + \frac{2}{3}\eta ^{\psi}\Big|_{\Nc\rightarrow 3}\, ,\\
\eta_{8}^{\mathrm{{SU(3)}}} 
 & = & 
  \Big[\eta_{1}^{A}+\eta_{2}^{A}
    \Big]_{\Nc\rightarrow3,c_{1}\rightarrow 3c_{1}/4}
  + \frac{2}{9}\eta ^{\psi}\Big|_{\Nc\rightarrow3,c_{1}\rightarrow
    c_{1}/3} 
  + \frac{4}{9}\eta ^{\psi}\Big|_{\Nc\rightarrow3,c_{1}\rightarrow
    4c_{1}/3}\, . 
\end{eqnarray}
The notation here serves as a recipe for replacing 
$\Nc$ and $c_{1}$,
defined in Eq. \eqref{eq:app_res_def_c1}, which appear on the
right-hand sides of Eqs. \eqref{eq:app_etaa2}, \eqref{eq:app_etab2} and
\eqref{eq:eta_quark2}.  Note that the replacement of $\Nc$ results
also in a modification of $z_{d}$, defined in Eq.
\eqref{eq:app_res_def_zd}. However, $c_{2}$, which appears in the
definition of $z_{d}$, remains unchanged for all gauge groups and
depends only on the dimension $d$.

\section{\label{sec:color}Color factors}

\setcounter{equation}{0}

In the following, we discuss the color factors $\tau_{i}^{A}$ and
$\tau_{i}^{\psi}$ which carry the information of the underlying
$SU(\Nc)$ gauge group.  First, we summarize the discussion of Ref.
\cite{Reuter:1997gx,Gies:2002af,Gies:2003ic} for the "gluonic" factors
$\tau_{i}^{A}$ appearing in the flow equation: Gauge group information
enters the flow of the coupling via color traces over products of
field strength tensors and gauge potentials.  For our calculation, it suffices 
to consider a pseudo-abelian background field
$\bar{A}$ which points into a constant color
direction $n^{a}$.  Therefore, the color traces reduce to
\begin{equation}
n^{a_{1}}n^{a_{2}}\dots n^{a_{2i}}\,\trc[T^{(a_{1}}T^{a_{2}}\dots
T^{a_{2i})}]\,,\label{eq:app_color_1} 
\end{equation}
where the parentheses at the color indices denote symmetrization.
These factors are not independent of the direction of $n^{a}$, but the
left-hand side of the flow equation is, since it is a function of the
$n^{a}$-independent quantity
$\frac{1}{4}F_{\mu\nu}^{a}F_{\mu\nu}^{a}$.  For this reason, we only
need that part of the symmetric invariant tensor $\trc[T^{(a_{1}}\dots
T^{a_{2i})}]$ which is proportional to the trivial one,
\begin{equation}
\trc[T^{(a_{1}}T^{a_{2}}\dots
T^{a_{2i})}]=\tau_{i}\,\delta_{(a_{1}a_{2}}\dots
\delta_{a_{2i-1}a_{2i})}+\dots\,. 
\label{eq:app_color_2}
\end{equation}
Here, we have neglected further nontrivial symmetric invariant tensors, since
they do not contribute to the flow of $\mathcal{W}_{k}(\theta)$, but
to that of other operators which do not belong to our truncation.
For the gauge group SU(2), there are no further symmetric invariant 
tensors in Eq. \eqref{eq:app_color_2},
implying 
\begin{equation}
\tau_{i}^{\text{SU(2)}}=2,\quad i=1,2,\dots\,\,.\label{eq:app_color_3}
\end{equation}
However, for higher gauge groups, the above mentioned complications
arise.  Therefore, we do not evaluate the $\tau_{i}^{A}$'s from Eq.
\eqref{eq:app_color_2} directly; instead, we use the fact that the
color unit vector $n^{a}$ can always be rotated into the Cartan
sub-algebra. Here, we choose the two color vectors $n^{a}$ which give
the extremal values for the whole trace of Eq.~\eqref{eq:app_color_1}.
For SU(3), these extremal choices are given by vectors $n^{a}$
pointing into the 3- and 8-direction in color space, respectively:
\begin{equation}
\tau_{i,3}^{A,\text{SU(3)}}=2+\frac{1}{4^{i-1}},
\quad\tau_{i,8}^{A,\text{SU(3)}}=3\,\left(\!\frac{3}{4}\!\right)^{i-1}\,.
\label{eq:app_color_4}
\end{equation}
Finally, we turn to the color factors $\tau _{j}^{\psi}$ of the quark sector. 
The above considerations also hold for the contributions of the flow
equation which arise from the fermionic part of our truncation
Eq.~\eqref{gentrunc}~and~\eqref{eq:quarktrunc1}. Taking into account
that quarks live in the fundamental representation and
choosing a color vector $n^{a}$ pointing into the 3- or 8-direction,
we obtain
\begin{equation}
\tau_{i,3}^{\psi,\text{SU(3)}}=2\,
\left(\!\frac{1}{4}\!\right)^{i},
\quad\tau_{i,8}^{\psi,\text{SU(3)}}=2
\,\left(\!\frac{1}{12}\!\right)^{i}+\left(\!\frac{1}{3}\!\right)^{i}
\quad i=1,2,\dots\,\, .
\label{eq:app_color_5}
\end{equation}
Again, all complications are absent for SU(2) and 
we find $\tau_{i}^{\psi,\text{SU(2)}}=\tau_{i,3}^{\psi,\text{SU(3)}}$.

The uncertainty introduced by the artificial $n^{a}$ dependence of the
color factors is the reason for the uncertainties of our results for
the critical temperature and the fixed point values in three and four
dimensions.

\section{Regulator dependence from the unstable mode}
\label{sec:regdep}
\setcounter{equation}{0}

In this section, we discuss the regulator dependence of the critical
temperature $T_{\text{cr}}$, arising from the details of projecting out the
unstable Nielsen-Olesen mode. As already explained in the main text,
removing the tachyonic part of the unstable mode corresponds to an
exact operation on the space of admissible stable background fields.
In the present context, it even suffices to remove only the thermal
excitations of the tachyonic part of the mode, since the imaginary
part arising from quantum fluctuations can easily be identified and
dropped. In the following, we take a less strict viewpoint and allow
for a smeared regularization of this mode in a whole class of
regulators. 

Since the true physical result will not depend on this part of the
regularization, we can identify the optimal (truncated) result with a
stationary point in the space of regulators, using the "principle of
minimum sensitivity", cf. \cite{Stevenson:1981vj}. 
\begin{figure}[t]
\includegraphics[%
  clip,
  scale=0.6]{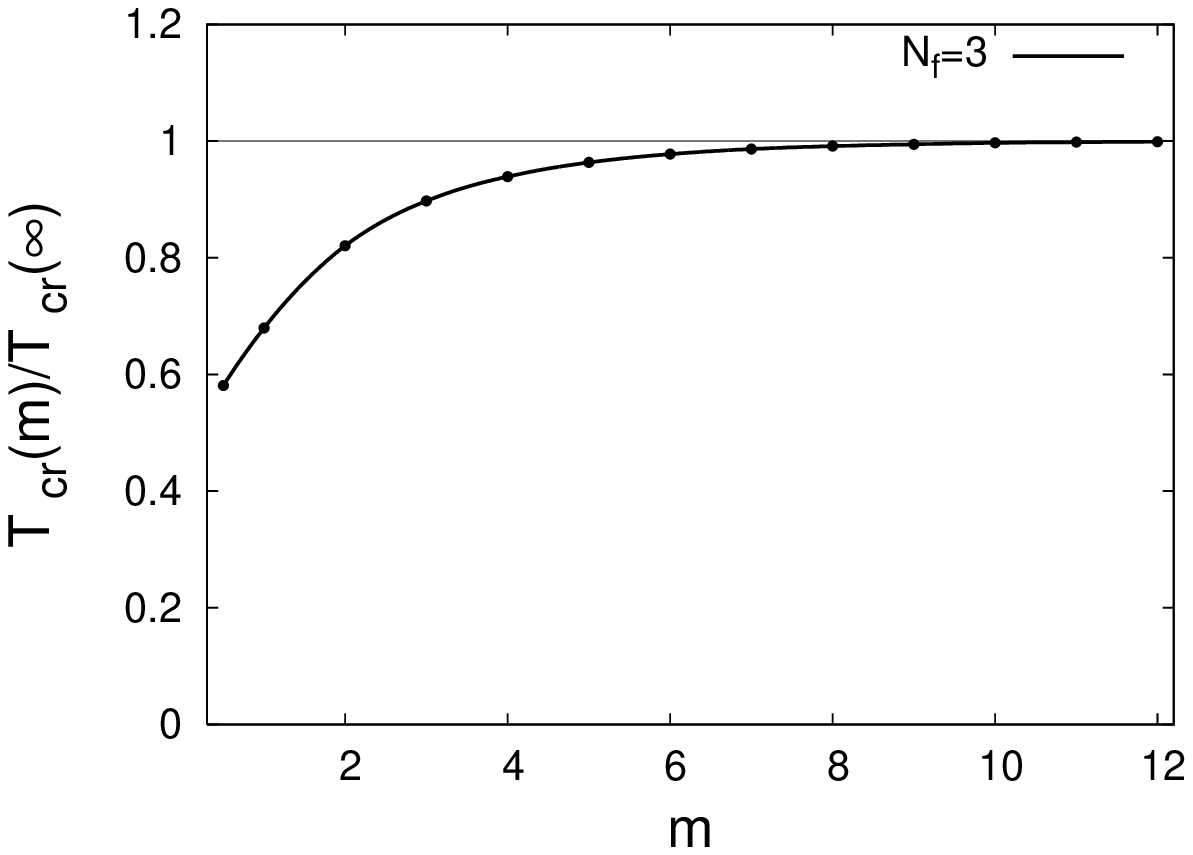}
\includegraphics[%
  clip,
  scale=0.6]{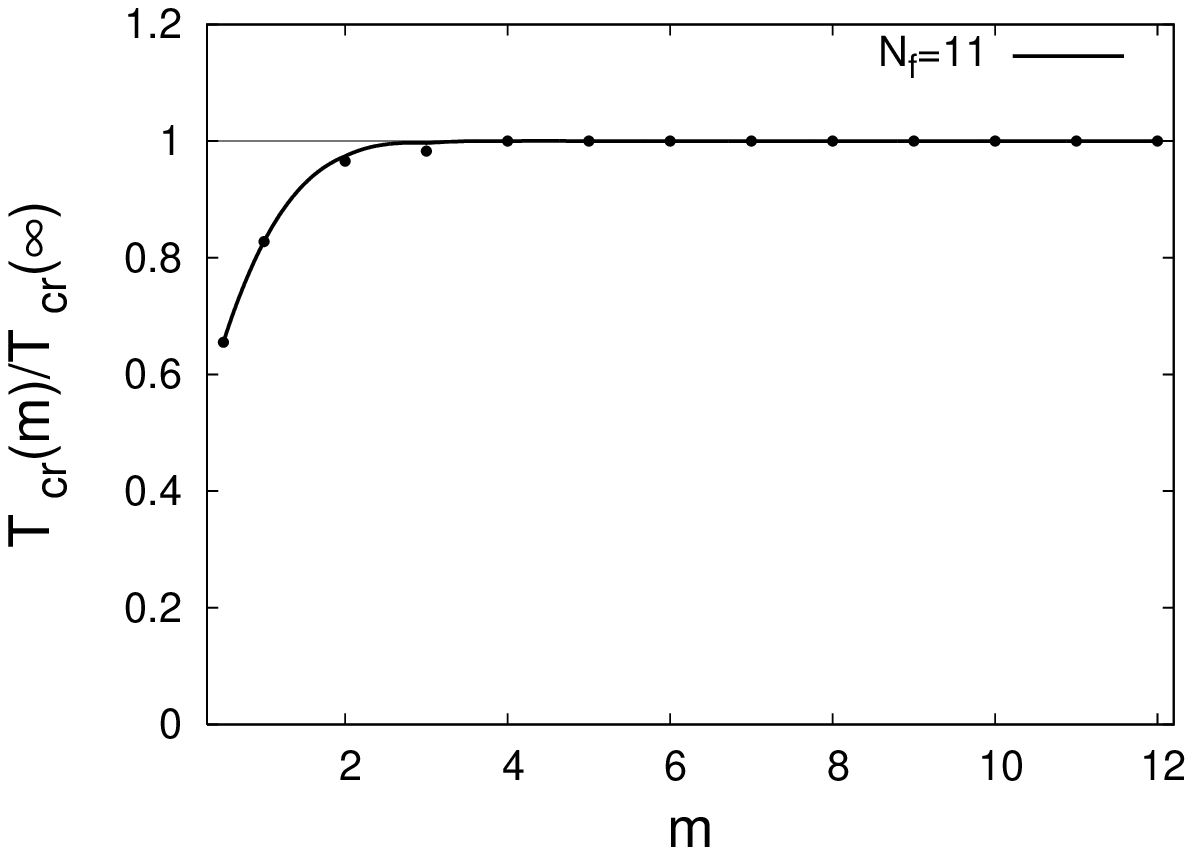}
\caption{Dependence of the critical temperature $T_c$ on the smeared
  regularization of the Nielsen-Olesen mode with $m$ labeling the
  regulator. The left and the right panel show the results for $\Nc$
  with $\Nf =3$ and $\Nf =11$ massless quark flavors, respectively.
  The limit $m\to\infty$ can be identified with the stationary point,
  and thus optimal regulator, in the class of considered regulators.
  This justifies constructively the procedure used in the main text
  which was derived from general considerations.}
\label{fig:plot_reg_dep} 
\end{figure}
In order to inhibit the thermal population of the Nielsen-Olesen mode
$E^{\text{NO}}$ at finite temperature, it suffices to regularize only 
the soft part (zero Matsubara frequency) of this mode as follows:
\begin{equation}
\frac{\mathrm{E}_{\text{soft}} ^{\mathrm{NO}} +
  R_{k}}{k^{2}}\quad\longrightarrow
\quad{\mathcal{P}}({\textstyle{\frac{T}{k}}})
+\frac{\mathrm{E} _{\text{soft}} ^{\mathrm{NO}} + R_{k}}{k^{2}}\,. 
\end{equation}
The function ${\mathcal{P}}({\textstyle{\frac{T}{k}}})$ has to 
satisfy the following constraints:
\begin{equation}
\lim _{T/k\rightarrow0}{\mathcal{P}}({\textstyle{\frac{T}{k}}})=0
\qquad\mathrm{and}\qquad\lim
_{T/k\rightarrow\infty}{\mathcal{P}}({\textstyle{\frac{T}{k}}})
\rightarrow\infty\,. 
\end{equation}
In the following, we choose 
\begin{equation}
{\mathcal{P}}({\textstyle{\frac{T}{k}}})\equiv{\mathcal{P}}_{m}({\textstyle{\frac{T}{k}}})=({\textstyle{\frac{T}{k}}})^{m}
\qquad\mathrm{with}\qquad m\,>\,0
\label{eq:no_reg}
\end{equation}
as a convenient example. As a regulator optimization condition, we
demand that $T_{\text{cr}}$ should be stationary with respect to a variation
of the optimal regulator function. Calculating $T_{\text{cr}}$ as a function
of the parameter $m$, the optimization condition for the regulator
function translates into
\begin{equation}
\frac{\partial T_{cr}}{\partial m}\Bigg| _{m=\bar{m}}\stackrel{!}{=}0\,.
\end{equation}
The solution $m=\bar{m}$ defines the desired optimized regulator.

As an example, we show $T_{cr}(m)/T_{cr}(\infty)$ as a function of $m$
for $\Nc=3$ with $\Nf=3$ and with $\Nf=11$ quark flavors in 
Fig.~\ref{fig:plot_reg_dep}.  We find that the optimized regulator is given
by $m\rightarrow\infty$ for all $\Nc$ and $\Nf$. This represents
an independent and constructive justification of the regularization
used in the main text, corresponding to the choice $m\to\infty$. 

\end{appendix}


\end{document}